\documentclass[vecphys]{svmult}

\usepackage{makeidx}         % allows index generation
\usepackage{graphicx}        % standard LaTeX graphics tool
\usepackage{multicol}        % used for the two-column index
\usepackage{amsmath, amssymb}
\makeindex             % used for the subject index
\newcommand{\II}{\mathbb{I}}
\newcommand{\ii}{\mathrm{i}}
\newcommand{\na}{\nabla}
\newcommand{\dd}{\mathrm{d}}
\newcommand{\pd}{\partial}
\newcommand{\hh}{\mathcal{H}}

\newcommand{\F}{\mathcal{F}}

\newcommand{\e}{\mathrm{e}}
\newcommand{\ket}[1]{\left|#1\right\rangle}
\newcommand{\bra}[1]{\left\langle #1\right|}
\newcommand{\bracket}[2]{\left\langle%
#1\left.\right|#2\right\rangle}

\newcommand{\tr}{\mathop{\mathrm{tr}}\nolimits}

\newcommand{\ft}[2]{{\textstyle\frac{#1}{#2}}}

\newcommand{\x}{\hat{x}}
\newcommand{\dop}{\hat{\delta}}

\begin{document}

\title*{Matrix Models}
\titlerunning{MM}
\author{Corneliu Sochichiu\inst{1,2,3}}
\authorrunning{C. Sochichiu}
\institute{
Max-Planck-Institut f\"ur Physik
(Werner-Heisenberg-Institut)
F\"ohringer Ring 6\\
80805 M\"unchen
\and
Institutul de Fizic\u a Aplicat\u
a A\c S, str. Academiei, nr. 5, Chi\c sin\u au\\ MD2028 MOLDOVA
\and
Bogoliubov Laboratory of Theoretical
Physics, Joint Institute for Nuclear Research\\ 141980 Dubna,
Moscow Reg., RUSSIA\\
\texttt{sochichi@mppmu.mpg.de}}
\maketitle

Matrix models and their connections to String Theory and
noncommutative geometry are discussed. Various types of matrix
models are reviewed. Most of interest are IKKT and BFSS models.
They are introduced as 0+0 and 1+0 dimensional reduction of
Yang--Mills model respectively. They are obtained via the
deformations of string/membrane worldsheet/worldvolume. Classical
solutions leading to noncommutative gauge models are considered.
% ----------------------------------------------------------
\section{Introduction}
\label{sec:Intro}

At the beginning let us define the topic of the present lectures.
As follows from the title, ``Matrix Models'' are theories in which
the fundamental variable is a matrix. The matrix variable can be a
just a constant or a function of time or even be defined as a
function over some space-time manifold. With this definition
almost any model existing in modern physics e.g. Yang--Mills
theory, theories of Gravity etc., will be a ``matrix theory''.
Therefore, when speaking on the matrix theory usually a simple
structure is assumed, e.g. when fundamental variables are constant
or at most time dependent. In the first case, the models of
\index{random matrices}random matrices, one has no time therefore
no dynamics. This is a statistical theory describing random matrix
distributions. These models are popular in many areas e.g. in the
context of description of integrable systems in QCD, or nuclear
systems as well as in the study of the lattice Dirac operators
(for a review see e.g.
\cite{Morozov:2005mz,Verbaarschot:2005rj,Brody:1981cx,%
Guhr:1997ve,Osborn:1998qb,Verbaarschot:1994qf,Shuryak:1992pi} and
references therein). The special case of interest for us are the
Yang--Mills type matrix models arising in String Theory such like
the \index{IKKT matrix model} \cite{Ishibashi:1996xs}.

Another case of of interest are so called \index{matrix
mechanics}matrix mechanics, i.e. theories of time-evolutive
matrices. These models along with the random matrix models are of
special interest in String Theory. Thus, the Yang--Mills type
matrix models appear to nonperturbatively describe collective
degrees of freedom in string theory called
\index{brane}\emph{branes}. Branes are extended objects on which
the ``normal'' fundamental strings can end. It was conjectured
that including the brane degrees of freedom in the
``conventional'' superstring theories leads to their unification
into the \emph{\index{M-theory}M-theory}, a model giving in its
different perturbative regimes all known superstring models. The
M-theory is believed to be related to the twelve-dimensional
membrane. In the light-cone frame it was conjectured to be
described by an Yang--Mills type matrix mechanics \index{BFSS
matrix model}(BFSS matrix model) \cite{Banks:1996vh}. As we will
see in the next section, this as well as the IKKT matrix model can
be obtained by quantization/deformation of, respectively, the
worldvolume of the membrane and the worldsheet of the string.

As it is by now clear, in this notes we are considering mainly
these two models, which sometimes are called ``matrix theories''
to underline their fundamental role in string theory.

The plan of these note is as follows. In the next section we give
the string motivation and introduce the matrix models as
dimensional reductions of supersymmetric Yang--Mills model. Next,
we consider Nambu--Goto description of the string and membrane and
show that the noncommutative deformation of the respectively,
worldsheet or worldvolume leads to IKKT or BFSS matrix models. In
the following section we analyze the classical solutions to these
matrix models and interpret them as noncommutative gauge models.
The fact that these models have a common description in terms of
the original matrix model allows one to establish the equivalence
relations among them.

\section{Matrix models of String Theory}
\label{sect:mmst}

\subsection{Branes and Matrices}
\label{ssect:BandM}

\index{brane}A breakthrough in the development of string theory,
\index{``the second string revolution''}``the second string
revolution'' happened when it was observed that in the dynamics of
fundamental string on has additional degrees of freedom
corresponding to the dynamics at the string ends
\cite{Polchinski:1995mt} (see \cite{Aharony:1999ti} for a review).

In the open string mode expansion the dynamics at the edge is
described by an Abelian gauge field (particle) (for a modern
introduction to string theory see e.g.
\cite{Polchinski:book,Kiritsis:1997hj}). The corresponding charge
of the end of the string is called \index{Chan--Patton
factor}\emph{Chan--Patton factor}. Allowing a superposition of
several, say $N$ such factors, which correspond to an
``$N$-valent'' string end, gives rise to a nonabelian U($N$) super
Yang--Mills gauge field in the effective lagrangian of the open
string. This is the so called nine-brane. As it was shown in
\cite{Polchinski:1995mt}  string theory allows brane
configurations of different other dimensions $p$, $0\leq p+1\leq
10$. Depending on the type of the string model they preserve parts
of supersymmetry.

So, descending down to the lower dimensional $p$-branes one gets
the $p+1$-dimensional reductions of the ten dimensional Super
Yang--Mills model.

It appears that out of all possibilities only two cases are
fundamental, namely, this of $p=0$ and $p=-1$. All other cases can
be obtained from either $p=-1$ or $p=0$ by condensation of $-1$-
or \index{$0$-branes}$0$-branes into higher dimensional objects.

\subsection{The IKKT matrix model family}
\index{IKKT matrix model}\index{random matrices} As it follows
from the space-time picture, the \index{$-1$-brane}$-1$ branes are
non-dynamical and, therefore, should be described by a random
matrix model which is the reduction of the 10d SYM down to zero
dimensions:
\begin{equation}\label{s-1}
  S_{-1}=-\frac{1}{4g^2}\tr[X_\mu,X_\nu]^2-
\tr\bar{\psi}\gamma^\mu[X_\mu,\psi],
\end{equation}
where $g$ is some coupling constant depending on SYM coupling $g_{\rm
  YM}$ and the volume of compactification. Matrices $X_\mu$,
$\mu=1,\dots, 10$ are Hermitian $N\times N$ matrices, $\psi$ is a
10d spinor which has $N\times N$ matrix index, $\gamma^\mu$ are 10d
Dirac $\gamma$-matrices.

From the 10d SYM the matrix model (\ref{s-1}) inherits the
following symmetries:
\begin{itemize}
\item Shifts:
  \begin{equation}
    X_\mu\to X_\mu+a_\mu\cdot \II,
  \end{equation}
where $a_\mu$ is a c-number.
\item SO(10) rotation symmetry
\begin{equation}\label{lorenz}
  X_\mu\to \Lambda_\mu{}^\nu X_\nu,
\end{equation}
where $\Lambda\in SO(10)$. This is the consequence of the
(euclideanized) Lorenz invariance of the ten dimensional SYM model.
\item SU(N) gauge symmetry
\begin{equation}\label{gauge}
  X_\mu\to U^{-1}X_\mu U,
\end{equation}
where $U\in SU(N)$, and this is the remnant of the SYM gauge symmetry
invariance.
\item Also one has left from the SYM model the
supersymmetry invariance:
\begin{align}\label{susy}
  \delta_1 X_\mu&=\bar{\epsilon}\gamma_\mu\psi,\\
  \delta_1 \psi&=[X_\mu,X_\nu]\gamma^{\mu\nu}\epsilon,\\
\intertext{as well as the second one which is simply the shift of the fermion,}
  \delta_2 X_\mu&=0,\\
  \delta_2\psi&=\eta,
\end{align}
where $\epsilon$ and $\eta$ are the supersymmetry transformation
parameters.
\end{itemize}

\begin{exercise}
 Find the relation between the coupling $g$ in (\ref{s-1}) on one side
 and SYM coupling $g_{\rm YM}$ and the size/geometry of
 compactification on the other side.
\end{exercise}
\underline{Hint}: Use an appropriate gauge fixing.

\begin{exercise}
 Show that (\ref{lorenz}--\ref{susy}) are indeed the symmetries of the
 action (\ref{s-1}).
\end{exercise}

The purely bosonic version of IKKT matrix model can be interpreted
as the algebraic version of much older \index{Eguchi--Kawai
model}Eguchi--Kawai model \cite{Eguchi:1982nm}. The last is
formulated in terms of SU(N) \emph{group} valued fields $U_\mu$
(in contrast to the algebra valued $X_\mu$). The action for the
Eguchi--Kawai model reads as,
\begin{equation}\label{ek}
  S_{\rm EK}=-\frac{1}{4g^2_{\rm EK}}\sum_{\mu,\nu}\tr(U_\mu U_\nu U^{-1}_\mu
  U^{-1}_\nu-\II).
\end{equation}
By the substitution, $U_\mu=\exp aX_\mu$, $g^2_{\rm EK}=g^2 a^{4-d}$
and taking the limit $a\to 0$ one formally comes to the bosonic part
of the IKKT action \eqref{s-1}.

\textbf{Note:} From the string interpretation we will discuss in
the next section it is worth to add an extra term to the IKKT
action \eqref{s-1} and, namely, the \index{chemical
potential}\index{$N$}chemical potential term,
\begin{equation}\label{ch-pot}
  \Delta S_{chem}=-\beta \tr \II,
\end{equation}
which ``controls'' the statistical behavior of $N$. In the
string/brane picture $\beta$ plays the role of the chemical
potential for the number of branes. This produces the relative
weights for the distributions with different $N$, which can not be
catched from the arguments we used to write down the action
\eqref{s-1}.

\subsection{The BFSS model family}
\index{BFSS matrix model} Let us consider another important model
which describes the dynamics of zero branes \cite{Banks:1996vh}.
Basic ingredients of this model are roughly the same as for the
previous one, the IKKT model, except that now the matrices depend
on time. The action for this model is the dimensional reduction of
the ten dimensional SYM model down to the only time dimension:
\begin{equation}\label{s0}
  S_{\rm BFSS}=\frac{1}{g_{\rm BFSS}}\int\dd t \tr\left\{
  \ft12(\na_0X_i)^2+\bar\psi\na_0\psi-\ft14[X_i,X_j]^2-
  \bar\psi\gamma_i[X_i,\psi]\right\},
\end{equation}
where, now, the index $i$ runs from one to nine.

The action \eqref{s0} describes the dynamics of zero branes in IIA
string theory, but it was also proposed as the action for the
M-theory membrane in the light-cone approach. As we are going to
see in the next section, this model along with the IKKT model can
be obtained by worldvolume quantization of the membrane action.

\index{BMN matrix model}Another known modification of this action
for the pp-wave background was proposed by
Berenstein--Maldacena--Nastase (BMN)
\cite{Berenstein:2002jq,Berenstein:2002zw,Berenstein:2002sa}. It
differs from the BFSS model additional terms which are introduced
in order to respect the pp-wave supersymmetry. The action of the
BMN matrix model reads:
\begin{multline}
  S_{\rm BMN}=\int \dd t \tr\left[\frac{1}{2(2R)}(\na_0X_i)^2+
  \bar\psi\na_0\psi\right.\\
  \left.+\frac{(2R)}{4}[X_i,X_j]^2-
   \ii (2R)\bar\psi\gamma^i[X_i,\psi]\right]+S_{mass},
\end{multline}
where $S_{mass}$ is given by
\begin{multline}
  S_{mass}=\int\dd
  t\tr\left[\frac{1}{2(2R)}\left(-\left(\frac{\mu}{3}\right)\sum_{i=1,2,3}X_i^2
  -\left(\frac{\mu}{6}\right)\sum_{i=4,\dots,9}X_i^2\right)\right.\\
  \left.-\frac{\mu}{4}\bar\psi\gamma_{123}\psi-
  \frac{\mu\ii}{3}\sum_{jkl=1,\dots,3}\epsilon_{ijk}X_iX_jX_k\right]
\end{multline}

The essential difference of this model from the standard BFSS one
is that due to the mass and the Chern-Simons terms this matrix
model allow stable vacuum solutions which can be interpreted as
\index{spherical branes} spherical branes (see e.g.
\cite{Valtancoli:2002rx,Sochichiu:2002ta}. Such vacuum
configurations can not exist in the original BFSS model.

\section{Matrix models from the noncommutativity}

In this section we show that the Matrix models which we introduced
in the previous section arise when one allows the worldsheet of
the string/worldvolume of the membrane to possess
\index{worldsheet deformation}\index{worldvolume deformation}
noncommutativity. It is interesting to note from the beginning
that the ``quantization'' of the string worldsheet leads to the
IKKT matrix model, while the space noncommutative membrane is
described by the BFSS model. Let us remind that the above matrix
models were introduced to describe, respectively, the $-1$- and
0-branes, while the string and the membrane are respectively 1-
and 2-brane objects. In the shed of the next section this can be
interpreted as deconstruction of the 1- and 2-branes into their
basic components, namely $-1$- and 0-brane objects.

\index{fermionic part}In this section we consider only the bosonic
parts. The extension to the fermionic part is not difficult, so
this is left to the reader as an exercise.

\subsection{Noncommutative string and the IKKT matrix model}

In trying to make the fundamental string noncommutative one
immediately meets the following problem: The noncommutativity
parameter is a dimensional parameter and, therefore, hardly
compatible with the worldsheet conformal symmetry which plays a
fundamental role in the string theory. Beyond this there is no
theoretical reason to think that the worldsheet of the fundamental
string should be noncommutative. On the other hand, the are other
string-like objects in the nonperturbative string theory:
D1-branes or D-strings. As it was realized, in the presence of the
constant nonzero Neveu-Schwarz $B$-field the brane can be
described by a noncommutative gauge models
\cite{Cheung:1998nr,Chu:1998qz,Chu:1999gi,Seiberg:1999vs}. Then,
in contrast to the fundamental string, it is natural to make the
D-string noncommutative.

Let us start with the Euclidean \index{Nambu--Goto
action}Nambu--Goto action for the string,
\begin{equation}\label{ng}
  S_{\rm NG}=T\int \dd^2\sigma \sqrt{\det_{ab}\pd_aX^\mu\pd_bX_\mu},
\end{equation}
where $T$ is the \index{D-string tension} D-string tension and
$X^\mu=X^\mu(\sigma)$ are the embedding coordinates. The
expression under the square root of the r.h.s. of \eqref{ng} can
equivalently rewritten as follows,
\begin{equation}
  \det\pd X\cdot\pd X=\ft12\Sigma^2,
\end{equation}
where,
\begin{equation}
  \Sigma^{\mu\nu}=\epsilon^{ab}\pd_aX^\mu\pd_bX^\nu,
\end{equation}
which is the induced the worldsheet volume form of the embedding
$X^\mu(\sigma)$.

The Nambu--Goto action then becomes:
\begin{equation}\label{bi}
  S_{\rm NG}=T\int\dd^2\sigma\, \sqrt{\ft12\Sigma^2}.
\end{equation}
This action is nonlinear and still quite complicate. A much simple
form can be obtained using the Polyakov trick. To illustrate the idea
of te trick which is widely used in the string theory consider first
the example of a particle.

\subsubsection{Polyakov's trick}

\index{Polyakov's trick} The relativistic particle is described by
the following reparametrization invariant action,
\begin{equation}\label{p}
  S_p=m\int\dd\tau\, \sqrt{\dot x^2},
\end{equation}
where $m$ is the mass and $x$ is the particle coordinate. The dynamics
of the particle \eqref{p} is equivalent, at least classically to one
described by the following action,
\begin{equation}\label{pp}
  S_{pp}=\int\dd\tau \left(\ft12e^{-1} \dot{x}^2+m^2 e\right).
\end{equation}
In this form one has a new variable $e$ which plays the role of the
line einbein function, or better to say of the one-dimensional volume
form.

To see the classical equivalence between \eqref{pp} and \eqref{p} one
should write down the equations of motion arising from the variation of
$e$,
\begin{equation}
  e^2=\frac{\dot{x}^2}{m^2},
\end{equation}
and use it to substitute $e$ in the action \eqref{pp} which should
give exactly \eqref{p}.
\begin{exercise}
  Show this!
\end{exercise}
As one can see, both actions \eqref{p} and \eqref{pp} are
reparametrization invariant, the difference being that the
Polyakov action \eqref{pp} is quadratic in the particle velocity
$\dot{x}$. This trick is widely used in the analysis of nonlinear
systems with gauge symmetry. In what follows we will apply it too.

Let us turn back to our string and the action \eqref{bi}. Applying
the Polyakov trick, one can rewrite the action \eqref{bi} in the
following (classically) equivalent form,
\begin{equation}\label{ngp}
  S_{\rm NGP}=\int\dd^2\sigma\,\left(\ft14\eta^{-1} \{X_\mu,X_\nu\}^2+
  \eta T^2\right),
\end{equation}
where $\eta$ is the string ``area'' density and we introduced the
Poisson bracket\index{Poisson bracket} notation,
\begin{equation}\label{pb}
  \{X,Y\}=\epsilon^{ab}\pd_a X\pd_b Y.
\end{equation}
It is not very hard to see that the bracket defined by \eqref{pb}
satisfies to all properties a Poisson bracket is supposed to satisfy.
\begin{exercise}
  Do it!
\end{exercise}

Let us note, that the Poisson bracket \eqref{pb} is not an
worldsheet reparametrization invariant quantity. Under the
reparametrizations $\sigma\mapsto\sigma'(\sigma)$ it transforms
like density rather than scalar the same way as $\eta$ is:
\begin{subequations}\label{rep}
\begin{align}
  \{X,Y\}&\mapsto
  \det\left(\frac{\pd\sigma'}{\pd\sigma}\right)\{X,Y\}'\\
  \eta&\mapsto
  \det\left(\frac{\pd\sigma'}{\pd\sigma}\right)\eta(\sigma')
\end{align}
\end{subequations}
Having two densities one can master a scalar,
\begin{equation}\label{scpb}
  \{X,Y\}_{s}=\eta^{-1}\{X,Y\},
\end{equation}
which is invariant. Actually, these two definitions coincide in
the gauge $\eta=1$, which in some cases may be possible only
locally. In terms of the scalar Poisson bracket the action is
rewritten in the form as follows
\begin{equation}
  S_{\rm NGP}=\int\dd^2\sigma\eta\,\left(\ft 14\{X_\mu,X_\nu\}^2+
  T^2\right),
\end{equation}
where $\dd^2\sigma\eta$ is the invariant worldsheet area form.

\subsubsection{``Quantization''}

Consider the naive quantization procedure we know from the quantum
mechanics. The classical mechanics is described by the canonical
classical Poisson bracket,
\begin{equation}
  \{p,q\}=1,
\end{equation}
and the quantization procedure consists, roughly speaking, in the
replacement of the canonical variables  $(p,q)$ by the operators
$\hat{p},\hat{q}$. At the same time the $-\ii \hbar\times$(Poisson
bracket) is replaced by the commutator of the corresponding
operators. In particular,
\begin{equation}\label{ha}
  \{p,q\}\mapsto[\hat{p},\hat{q}]=-\ii\hbar.
\end{equation}
Afterwards, main task consists in finding the irreducible
representation(s) of the obtained algebra\footnote{In fact, the enveloping
  algebra rather the Lie algebra itself.}. From the undergraduate course
of quantum mechanics we know that there are many unitary
equivalent ways to do this, e.g. the \index{oscillator
basis}oscillator basis representation is a good choice.

Under the quantization procedure functions on the phase space are
replaced by operators acting on the irreducible representation
space of the algebra \eqref{ha}. For these functions and operators
one have the correspondence between the tracing and the
integration over the phase space with the Liouville measure
\begin{equation}
  \int\frac{\dd p\dd q}{2\pi\hbar} \dots\mapsto \tr \dots
\end{equation}

\index{worldsheet quantization}Let us turn to our string model. As
in the case of quantum mechanics, under the quantization we mean
the replacing the fundamental worldsheet variables $\sigma^1$ and
$\sigma^2$ by corresponding operators: $\hat\sigma^1$ and
$\hat\sigma^2$,  such that the \emph{invariant} Poisson bracket is
replaced by the commutator according to the rule:
\begin{equation}\label{dirac1}
  \{\cdot ,\cdot\}_{\rm PB}=\ii/\theta [\cdot,\cdot],
\end{equation}
where $\theta$ is the deformation parameter (noncommutativity).
The worldsheet functions are  replaced by the operators on the
Hilbert space on which $\hat\sigma^a$ act irreducibly. As we  have
two forms of the Poisson bracket the question is wether one should
use the density form of the Poisson bracket \eqref{pb} or the
invariant form \eqref{scpb}? The correct choice is the invariant
form \eqref{scpb}. This is imposed by the fact that the operator
commutator is invariant with respect of the choice of basic
operator set (in our case it is given by operators
$\hat\sigma^a$).

Let us note that with the choice of invariant Poisson bracket in
\eqref{dirac1} the operators $\hat\sigma^a$, generally, do not have
standard Heisenberg commutation relations. Rather than that, they
commute to a nontrivial operator,
\begin{equation}\label{s-comm}
  [\hat\sigma^1,\hat\sigma^2]=\ii\theta\widehat{\eta^{-1}},
\end{equation}
where the operator $\widehat{\eta^{-1}}$ corresponds to the inverse
density of the string worldsheet area (i.e. its classical limit gives
this density). At the same time the trace in the quantum case
corresponds to the worldsheet integration with the invariant measure
\begin{equation}\label{int2tr}
  \int\dd^2\sigma\eta\,[\dots] \mapsto 2\pi\theta \tr[\dots].
\end{equation}

Having the ``quantization rules''  \eqref{dirac1} and
\eqref{int2tr} one is able to write down the noncommutative analog
of the Nambu--Goto--Polyakov string \index{IKKT matrix
model}action \eqref{ngp}. It looks as follows,
\begin{equation}\label{mmncs}
  S=\alpha\tr\ft14[X_\mu,X_\nu]^2+\beta\tr\II,
\end{equation}
where $\alpha$ and $\beta$ are the couplings of the matrix model.
In terms of the string and the deformation parameters they read,
\begin{align}
  \alpha &=\frac{2\pi}{\theta}, \\
  \beta  &=\frac{2\pi T^2}{\theta}.
\end{align}

After the identification of couplings the model \eqref{mmncs} is
identic with the IKKT model \eqref{s-1}. As a bonus we have
obtained the chemical potential \eqref{ch-pot}. As we see from the
construction, the dimensionality of matrices depend on the
irreducibility representation of the noncommutative algebra. As
one can expect from what is familiar in quantum mechanics, the
compact worldsheets should lead to \emph{finite-dimensional}
representations and thus are described, respectively, by matrices
of finite dimensions. There is no exact equivalence between the
worldsheet geometry and the matrix description. However, the
consistency requires that one should recover the worldsheet
geometry in the semi-classical limit ($\theta\to 0$).

Another interesting remark is that in this picture the Heisenberg
operator basis correspond to the worldsheet parametrization for
which $\eta$ is constant. as it is well known such parametrization
can exist globally only for the topologically trivial worldsheets.
On the other hand, in the algebra of operators acting on a
separable infinite dimensional Hilbert space one can always find a
Heisenberg operator basis.

\subsubsection{Example I: Torus}

\index{noncommutative torus}To illustrate the above consider the
example of quantization of toric worldsheet. The torus can be
described by one complex modulus (or two real moduli). We are not
interested here in the possible form of the toric metric, so we
can choose the parametrization of the torus for which $\eta=1$ and
the flat worldsheet coordinates span the range
\begin{equation}\label{sigma12}
  0\leq\sigma^1<l_1,\qquad 0\leq\sigma^2<l_2.
\end{equation}

The first problem arises when one tries to quantize variables with the
range \eqref{sigma12}. In spite of the fact that the (invariant)
Poisson bracket is canonical the operators $\hat\sigma^{1,2}$ can not
satisfy the Heisenberg algebra,
\begin{equation}
  [\hat\sigma^1,\hat\sigma^2]=\ii\theta,
\end{equation}
and have bounded values like in \eqref{sigma12} at the same time.
\begin{exercise}
  Prove this!
\end{exercise}

To conciliate the compactness and noncommutativity one should use the
compact coordinates $U_a$ instead,
\begin{equation}
  U_a=\exp 2\pi\ii\hat\sigma_a/l_a,\qquad a=1,2.
\end{equation}
The compact coordinates $U_a$ satisfy the following (Weyl)
commutation relations\index{noncommutative torus}
\begin{equation}
  U_1 U_2=q U_2 U_1,
\end{equation}
where $q$ is the toric deformation parameter,
\begin{equation}
  q=\e^{2\pi^2\ii\theta/l_1l_2}.
\end{equation}

If $q^N=1$ for some $N\in\mathbb{Z}_+$, then $U_a$ generate an
irreducible representation of dimension $N$. In this case an arbitrary
$N\times N$ matrix $\mathcal{M}$ can be expanded in powers of $U_a$, e.g.
\begin{equation}\label{expansion}
  \mathcal{M}=\sum_{m,n=0}^{N-1}M_{mn}U_1^mU_2^n.
\end{equation}

Expansion \eqref{expansion} is in terms of monomials in $U_1$ and
$U_2$ ordered in such a way that all $U_1$'s are to the left of all
$U_2$ one can alternatively use the Weyl functions $W_{mn}$ defined as
\begin{equation}\label{weyl1}
  W_{mn}=\exp \left(2\pi\ii m\hat\sigma_1/l_1+2\pi\ii n\hat\sigma_2/l_2\right),
\end{equation}
which differs from the product $U_1^m U_2^n$ by a polynomial of
lower degree, but is symmetrized in $\hat\sigma^1$ and
$\hat\sigma^2$. Using this expansion in terms of the Weyl
functions leads one to the description of matrices in terms of the
\index{Weyl symbol}\emph{Weyl symbols} --- ordinary functions
subject to the \index{star product}\emph{star product} algebra.
Weyl symbols as well as the star product algebras we are going to
consider in the next sections.

As a result we have that quantization of the torus surface leads
to the description in terms of $N\times N$ matrices where  the
dimensionality $N$ of the matrices depends on the torus moduli.

\subsubsection{Example II: Fuzzy sphere}

\index{fuzzy sphere}Another case of interest is the deformation of
the spherical string worldsheet. On the sphere there is no global
flat parametrization with $\eta=1$. It is convenient to represent
the two-sphere worldsheet parameters embedded into the
three-dimensional Euclidean space:
\begin{equation}
  \sigma_1^2+\sigma_2^2+\sigma_3^2=1,
\end{equation}
with the induced metric and volume form $\eta$. The (invariant)
Poisson bracket is given by the following expression\footnote{We
drop out the subscript of the invariant Poisson bracket since it
creates no confusion while it is the only used from now on.}:
\begin{equation}\label{liePB}
  \{\sigma_i,\sigma_j\}=(1/r)\epsilon_{ijk}\sigma_k.
\end{equation}

Quantization of the Poisson algebra \eqref{liePB} leads to the su(2)
Lie algebra commutator,
\begin{equation}
  [\hat\sigma_i,\hat\sigma_j]=\ii(\theta/r)\epsilon_{ijk}\hat\sigma_k,
\end{equation}
whose unitary irreducible representations are the well known
representations of the su(2) algebra. They are parameterized by
the spin of the representation $J$. The dimensionality of such
representation is $N=2J+1$. The two dimensional parameters: the
radius of the sphere and the noncommutativity parameter are not
independent. They satisfy instead,
\begin{equation}
  r^4=\theta^2J(J+1).
\end{equation}

Again, arbitrary $(2J+1)\times(2J+1)$ matrix can be expanded in terms
of symmetrized monomials in $\sigma_i$ --- \emph{noncommutative
spherical harmonics}, which are the spherical analogues of the Weyl
functions.

Turning back to the action one get exactly the same model as in
the previous example with $N=2J+1$. As a result we get that
independently from which geometry one starts one gets basically
the same deformed description. The only meaningful parameter is
the dimensionality of the matrix and it depends only on the
worldsheet area. This is a manifestation of the universality of
the matrix description which we plan to explore in the next
sections.

\subsection{Noncommutative membrane and the BFSS matrix model}

\index{BFSS matrix model}Let us consider slightly more complicate
example, namely that of the membrane. For the membrane one can
write a \index{membrane}\index{Nambu--Goto action}Nambu--Goto
action too,
\begin{equation}
  S_{NG}=T_m\int_{\Sigma_3}\dd^3\sigma\,\sqrt{-\det\pd_a X^\mu\pd_b X_\mu},
\end{equation}
where $T_m$ is the membrane tension and $X$ are the membrane embedding
functions.

In the case when the topology of the worldvolume $\Sigma_3$ is of
the type $\Sigma_3=I\times \mathcal{M}_2$, where $\mathbb{R}^1$ is
the time interval $I=[0,t_0]$ and $\mathcal{M}_2$ is a two
dimensional manifold, one has the freedom to choose the worldsheet
parameters $\sigma^i$, $i=1,2,3$ in such a way that the time like
tangential will be always orthogonal to the space-like tangential,
\begin{equation}\label{constr1}
  \pd_0X^\mu\pd_a X_\mu=0.
\end{equation}
In this case the Nambu--Goto action takes the following form
\begin{equation}
  S_{NG}=T_m\int\dd\tau \dd^2\sigma\,\sqrt{\ft12\dot{X}^2\Sigma_{\mu\nu}^2},
\end{equation}
where,
\begin{equation}
  \Sigma_{\mu\nu}=\epsilon_{ab}\pd_a X_\mu\pd_b X_\nu.
\end{equation}

In the complete analogy to the case of the string let us rewrite
the Nambu--Goto action in the \index{Polyakov's trick}Polyakov
form,
\begin{equation}
  S_{\rm NGP}=\int\dd^3\sigma \eta
  \left[\ft{T_m^2}{2}\dot{X}^2+\ft14\{X_\mu,X_\nu\}^2 \right],
\end{equation}
where the (invariant) \index{Poisson bracket}Poisson bracket is
defined as
\begin{equation}
  \{X,Y\}=\eta^{-1}\epsilon_{ab}\pd_X\pd_b Y.
\end{equation}
Since we partially fixed the reparametrization gauge invariance by
choosing the time direction we have the \index{reparametrization
invariance}constraint \eqref{constr1}. This leads to the following
constraint,
\begin{equation}\label{constr2}
  \{\dot{X}^\mu,X_\mu\}=0.
\end{equation}

Now, straightforwardly repeating the arguments of the previous
subsection one can write down the matrix model action. In the
\index{BFSS matrix model}present case the action takes the
following form:
\begin{equation}\label{m}
  S_{m}=\int\dd t\left(\beta \tr\ft12 \dot{X}^2+\alpha\tr\ft14[X_\mu,X_\nu]^2\right),
\end{equation}
where $\beta= 2\pi T^2/\theta$ and $\alpha=2\pi/\theta$, respectively.
The action \eqref{m} should be supplemented with the following
constraint:
\begin{equation}\label{constr3}
  [\dot{X}_\mu,X_\mu]=0.
\end{equation}

The constraint \eqref{constr3} can be added to the action
\eqref{m} with the Lagrange multiplier $A_0$. In this case the
action acquires the following form:
\begin{equation}\label{bfss-a}
  S_{gi}=\int\dd t\left(\beta\tr\ft12 (\na_0X_\mu)^2
  +\alpha\tr\ft14[X_\mu,X_\nu]^2\right),
\end{equation}
which is identic (upto definition of parameters $\alpha$ and
$\beta$) to the bosonic part of the BFSS action \eqref{s0}. By the
redefinition of the matrix fields and rescaling of the time one
can eliminate the constants $\alpha$ and $\beta$, so in what
follows we can put both to unity.

\index{fermionic part}So far we have considered only the bosonic
parts of the membrane. Including the fermions (when they exist)
introduces no conceptual changes. Therefore, derivation of the
fermionic parts of the IKKT and BFSS matrix model description of
the string and membrane is entirely left to the reader.
\begin{exercise}
  Derive the fermionic part of both matrix models starting from
  the superstring/supermembrane.
\end{exercise}

\section{Equations of motion. Classical solutions}

In this section we consider two types of theories, namely the
string and the membrane in the Nambu--Goto-Polyakov form and the
corresponding matrix models. One can write down equations of
motion and try to find out some simple classical solutions in
order to compare these cases among each other.

The static equations of motion in the membrane case coincide with the
string equations of motion. Therefore, it is enough to consider only
the last case: Any solution in the IKKT model has also the interpretation
as a classical vacuum of the BFSS theory.

\subsection{Equations of motion before deformation:
  Nambu--Goto--Polyakov string}

\index{Nambu--Goto--Polyakov equations of motion}Consider first
the equations of motion corresponding to the Nambu--Goto--Polyakov
string \eqref{ngp} in the form one gets just before the
deformation procedure.

Variation of $X_\nu$ produces the following equations,
\begin{subequations}\label{eom-cl}
\begin{equation}\label{ngp-x-eom}
  \{X_\mu,\eta^{-1} \{X_\mu,X_\nu\}\}=0,
\end{equation}
while the variation of $\eta$ produces the constraint
\begin{equation}\label{ngp-eta-eom}
  \eta^2=\frac14\frac{\{X_\mu,X_\nu\}^2}{T^2}.
\end{equation}
\end{subequations}
(As in the Polyakov particle case the last equation can be used to
eliminate $\eta$ from the action \eqref{ngp} in order to get the
original Nambu--Goto action \eqref{ng}.)

The equations of motion \eqref{eom-cl} posses a large symmetry
related to the reparametrization invariance \eqref{rep}. In order
to find some solutions it is useful (but not necessary!) to fix
this gauge invariance. As the use of the model is to describe
branes, one may be interested in solutions corresponding to
infinitely extended branes, which have the topology of
$\mathbb{R}^2$. In this, simplest case one can impose the gauge
$\eta=1/4T^2$. Then, the equations of motion \eqref{eom-cl} are
reduced to
\begin{equation}\label{eom-gauged}
  \{X_\mu,\{X_\mu,X_\nu\}\}=0,\qquad \{X_\mu,X_\nu\}^2=1.
\end{equation}

In the case of two dimensions ($\mu,\nu=1,2$), one can find even
the generic solution. It is given by an arbitrary canonical
transformation $X_{1,2}=X_{1,2}(\sigma_1,\sigma_2)$. This is easy
to see if to observe that the second equation in
\eqref{eom-gauged} requires that the $XX$ Poisson bracket must be
a canonical one. The first equation is then satisfied
automatically. One can also see that all the arbitrariness in the
solution is due to the remnant of the reparametrization invariance
which is given by the \index{area
  preserving diffeomorphisms}\emph{area
  preserving diffeomorphisms}. This situation is similar to one
met in the case of two dimensional gauge theories where there are
no physical degrees of freedom left to the gauge fields beyond the
gauge arbitrariness. As we will see later, this similarity is not
accidental, in some sense the above matrix model is indeed a
two-dimensional gauge theory.

The situation is different in more than two dimensions. In this
case we are not able to write down the generic solution, but one
can find a significant particular one. The simplest solutions of
\eqref{eom-gauged} can be obtained by just lifting up the
two-dimensional ones to higher dimensions. In particular, one has
the following solution
\begin{equation}\label{br-sol}
  X_1=\sigma_1 ,\quad X_2=\sigma_2,\quad X_i=0,\quad
  i=3,\dots,10.
\end{equation}
It is not difficult to check that the solution \eqref{br-sol} satisfy
to both equations \eqref{eom-gauged}. The physical meaning of this
solution is an infinite Euclidean brane extended in the plane (1,2).

One can see, that by the nature of the model in which fields
$X_\mu$ are functions of a two dimensional parameter the solutions
to the equations of motion are forced always to describe two
dimensional surfaces i.e. single brane configurations. One can go
slightly beyond this limitation allowing $X$'s to be multivalent
functions of $\sigma$'s. In this case one is able to describe a
certain set of multibrane systems, each sheet of $X$ corresponding
to an individual brane. This situation in application to spherical
branes was analyzed in more details in \cite{Sochichiu:2002ta}.

Another question one may ask is whether one can find solutions
describing a compact worldsheet. We are not going to give any
proof of the fact that such type of solutions do not exist. Rather
we consider a simple example of a cylindrical configuration and
show that thew equations of motion are not satisfied by it. An
infinite cylinder as an extremal case of the torus can be given by
the following parametric description:
\begin{equation}\label{cyl}
  X_1=\sin \sigma_1,\quad X_2=\cos\sigma_1,\quad X_3=\sigma_2.
\end{equation}
The eq. \eqref{cyl} describes a cylinder obtained from moving the
circle in the plane (1,2) along the axe 3. The parametrization
\eqref{cyl} satisfy the constraint \eqref{constr1}, therefore to see
wether such surface is a classically stable it is enough to check the
the first equation of \eqref{eom-gauged}. The explicit evaluation of
the equations of motion gives
\begin{align}
  \{X_\mu,\{X_\mu,X_1\}\}&=-X_1\neq 0, \\
  \{X_\mu,\{X_\mu,X_2\}\}&=-X_2\neq 0, \\
  \{X_\mu,\{X_\mu,X_0\}\}&=0.
\end{align}
As one see, only the equation of motion for the third noncompact
direction is satisfied. Other equations can be satisfied if one
modifies the action of the model by adding mass terms for e.g.
\index{mass term}$X_1$ and $X_2$:
\begin{equation}
  S\to S+m^2 (X_1^2+X_2^2).
\end{equation}

\begin{exercise}
  Modify the classical action in a way to allow the spherical brane
  solutions. Worldsheet quantize this model and compare it to the
  BMN  matrix model.
\end{exercise}

Another interesting type of solutions is given by singular
configurations with trivial Poisson bracket,
\begin{equation}\label{zero}
  \{X_\mu,X_\nu\}=0.
\end{equation}
Obviously, these configurations satisfy the equations of motion.
This solution corresponds to an arbitrary open or closed smooth
one-dimensional line embedded in $\mathbb{R}^D$. The problem
appears when one tries to make this type of solution to satisfy
the constraint \eqref{constr1} arising from the gauge fixing
$\eta^2=1/4T^2$. This configuration, however is still an
acceptable solution before the gauge fixing. The degeneracy of the
two dimensional surface into the line results into the degeneracy
of the two-dimensional surface reparametrization symmetry into the
subgroup of the line reparametrizations. This means in particular
that $\eta^2=1/4T^2$ is not an acceptable gauge condition in this
point, one must impose $\eta=0$ instead.

Let us now turn to the noncommutative case and see how the situation
is changed there.

\subsection{Equations of motion after deformation: \\ IKKT/BFSS matrix
  models}

After quantization of the worldsheet/worldvolume we are left with
no Polyakov auxiliary field $\eta$. The role of this field in the
noncommutative theory is played by the choice of the
representation. As most cases we can not smoothly variate the
representation, we have no equations of motion corresponding to
this parameter. So, we are left with only equations of motion
corresponding to the variation of $X$'s. For the IKKT model these
equations read
\begin{equation}\label{ikkt-eom}
  [X_\mu,[X_\mu,X_\nu]]=0,
\end{equation}
while for the BFSS model the variation of $X$ leads to the
following dynamical equations,
\begin{equation}\label{bfss-eom}
  \ddot{X}_\mu+[X_\mu,[X_\mu,X_\nu]]=0,
\end{equation}
where we also put the brane tension to unity: $T=1$. If one is
interested in only the static solutions ($\dot{X}=0$) to the BFSS equations of
motion, then the equation  \eqref{bfss-eom} is reduced down to the
IKKT equation of motion. Therefore, in what follows we consider only
the last one.

By the first look at the equation \eqref{ikkt-eom} it is clear that
one can generalize the string soluiton \eqref{br-sol} from the
commutative case. Namely, one can check that the configuration
\begin{equation}\label{ha-sol}
  X_1=\hat\sigma_1,\qquad X_2=\hat\sigma_2,\qquad X_i=0,\quad
  i=3,\dots, D,
\end{equation}
satisfy the equations of motion \eqref{ikkt-eom}. By the analogy with
the commutative case we can say that this configuration describes
either Euclidean D-string (IKKT) or a static membrane (BFSS). The
solution \eqref{ha-sol} corresponds to the Heisenberg algebra
\begin{equation}\label{sol1}
  [X_1,X_2]=1,
\end{equation}
which allows only the infinite-dimensional representation. The
value of $X$ are not bounded, therefore this solution corresponds
to a noncompact brane.

What is the role of the $\eta$-constraint here? The algebra
\eqref{sol1} does not completely specify the solution unless the
nature of its representation is also given. In particular, the
algebra of $\hat{\sigma}$'s can be irreducibly represented on the
whole Hilbert space. In the semiclassical limit this can be seen
to correspond to the constraint of the previous subsection.

As we discussed in the case of commutative string, any solution to
the equations of motion describes a two dimensional surface and,
therefore, has the Poisson bracket of the rank (in indices $\mu$
and $\nu$) two or zero. In contrast to this, in the noncommutative
case one may have solutions with an arbitrary even rank between
zero and $D$. Indeed, consider a configuration,
\begin{equation}\label{heis}
  X_a=p_a,\quad a=1,\dots, p+1,\qquad X_i=0,\quad i=p+2,\dots,D,
\end{equation}
such that
\begin{equation}\label{d-ha}
  [p_a,p_b]=\ii B_{ab},\qquad \det B\neq 0,
\end{equation}
where $B$ is the matrix with c-number entries $B_{ab}$. Such set
of operators always exists if the Hilbert space is infinite
dimensional separable. The set of operators $p_a$ generate a
Heisenberg algebra. Interesting cases are when the Heisenberg
algebra \eqref{d-ha} is represented irreducibly on the Hilbert
space of the model, or when this irreducible representation is
$n$-tuple degenerate. Analysis of these cases we will do in the
next sections.

%%%%%%%%%%%%%%%%%%%%%%%%%%%%%%%%%%%%%%%%%%%%%%%%%%%%%%%%%%%%%%%%%%%%%
\index{compact branes}How about the compact branes? As we have
already discussed in the previous section, the compact worldsheet
solution corresponds to finite dimensional matrices $X_\mu$. As it
appears for such matrices the only solution to the equation of
motion which exists is one with the trivial commutator,
\begin{equation}
  [X_\mu,X_\nu]=0.
\end{equation}
To prove this fact, suppose we find such a solution with
$B_{\mu\nu}=[X^{(0)}_\mu,X^{(0)}_\nu]\neq 0$ and satisfying the
equations of motion
\eqref{ikkt-eom}. The IKKT action (BFSS energy) computed on such a
solution is
\begin{equation}
  S(X)=-\ft14B_{\mu\nu}^2\tr\II \neq 0.
\end{equation}
Since this is a solution to the equations of motion the variation
of the action should vanish on the solution,
\begin{equation}
  \delta S=\tr\frac{\delta S}{\delta X_\mu}(X^{(0)})\delta X_\mu=0,
  \qquad \text{for
  }\forall\, \delta X_i,
\end{equation}
which is not the case: Take $\delta X_\mu=\epsilon X^{(0)}_\mu$ to find out
that $\delta S|_{X^{(0)}}\neq 0$. So there are no solutions with
nontrivial commutator for the finite dimensional matrix space.

Consider now the extremal case of singular solutions with vanishing
commutators,
\begin{equation}\label{zero-m}
  [X_\mu,X_\nu]=0.
\end{equation}
Obviously, from the equation \eqref{zero-m} automatically follows
that the equations are satisfied too. This solution exists in both
finite as well as infinite-dimensional cases. Since the
commutativity of $X_\mu$'s allows their simultaneous
diagonalization
\begin{equation}\label{diag}
  X_\mu =
  \begin{pmatrix}
    x_1^\mu & & &\\
    & x_2^\mu & & \\
    & &  \ddots &
  \end{pmatrix},
\end{equation}
this means that the branes which are described by the matrix models
are localized $x^\mu_k$ being the coordinates of the $k$-th brane.

\subsubsection{The symmetry of the solutions}

The various types of solutions have different symmetry properties.
Thus, the solution of the type \eqref{heis} with the algebra of
$p_a$'s irreducibly represented over the Hilbert space of the
model has no internal symmetries. Indeed, by the Schurr's lemma
any operator commuting with all $p_a$ is proportional to the
identity. In the case when the representation is $n$-tuple
degenerate one has an U$(n)$ symmetry mixing the representations.
The degenerate case \eqref{zero-m}, when $B_{\mu\nu}=0$ give rise
to some symmetries too. Indeed, an arbitrary diagonal matrix
commute with all $X_\mu$ given by \eqref{diag}. If no two branes
are in the same place: $x_m^\mu\neq x_n^\mu$ for any $m\neq n$,
then the configuration breaks the U$(N)$ symmetry group (in the
finite-dimensional case) down to the the Abelian subgroup
U(1)$^N$.

%%%%%%%%%%%%%%%%%%%%%%%%%%%%%%%%%%%%%%%%%%%%%%%%%%%%%%%%%%%%%%%%%%%%%
\section{From the Matrix Theory to Noncommutative Yang--Mills}

This and the following section is mainly based on the papers
\cite{Sochichiu:2000ud,Sochichiu:2000fs,Sochichiu:2000bg,Sochichiu:2000kr,%
Sochichiu:2000kz,Kiritsis:2002py}, the reader is also referred to
the lecture notes \cite{Sochichiu:2002jh} and references therein.

The main idea is to use the solutions from the previous section both
as classical vacua, such that arbitrary matrix configuration is
regarded as a perturbation of this vacuum configuration, and as a
basic set of operators in terms of which the above perturbations are
expanded. Now follow the details.

\subsection{Zero commutator case: gauge group of diffeomorphisms}

Consider first the case of the solution with the vanishing
commutator \eqref{zero-m}. We are interested in configurations in
which the branes form a $p$-dimensional lattice. Using the
rotational symmetry of the model, one can choose this lattice to
be extended in the dimensions $1,\dots, p$:
\begin{equation}\label{bg}
  X_a\equiv p_a,\quad a+1,\dots,p;\qquad X_I=0, \quad I=p+1,\dots D.
\end{equation}

Then an \emph{arbitrary} configuration can be represented as
\begin{equation}\label{pert-zero-m}
  X_a=p_a+A_a,\qquad X_I=\Phi_I.
\end{equation}

Let us take the limit $N\to\infty$ and take such a distribution of
the branes in which they form an infinite regular $p$-dimensional
lattice:
\begin{equation}
  p_a\to \lambda n_a, \qquad n_a\in \mathbb{Z},
\end{equation}
such that the Hilbert space can be split in the product of $p$
infinite-dimensional subspaces $\hh_a$
\begin{equation}\label{hs-spl}
  \hh=\otimes_{a=1}^p \hh_a,
\end{equation}
such that each eigenvalue $\lambda n_a$ is non-degenerate in
$\hh_a$. In this case the operators $p_a$ can be regarded as
($-\ii$ times) partial derivatives on a $p$-dimensional torus of
the size $1/\lambda$,
\begin{equation}\label{p_a}
  p_a=-\ii \pd_a.
\end{equation}

Now let us turn to the perturbation of the vacuum configuration
\eqref{pert-zero-m} and try to write it in terms of operators
$p_a$. Since the algebra of $p_a$'s is commutative, they alone
fail to generate an irreducible representation in terms of which
one can expand an arbitrary operator acting on the Hilbert space
$\hh$. One must instead supplement this set with with $p$ other
operators $x^a$, which together with $p_a$ form a Heisenberg
algebra irreducibly represented on $\hh$,
\begin{equation}\label{0-m-heis}
  [x^a,x^b]=0,\qquad [p_a,x^b]=-\ii \delta_a{}^b.
\end{equation}

From the algebra \eqref{0-m-heis} follows that the operators $x^a$
have a continuous spectrum which is bounded: $-\pi/\lambda\leq
x^a< \pi/\lambda $. This precisely means that $x^a$ are operators
of coordinates on the $p$-dimensional torus. Then, an arbitrary
matrix $X$ can be represented as a an operator function of the
operators $p_a$ and $x^a$,
\[
  X=\hat{X}(\hat{p},\hat{x}).
\]
In the ``$x$-picture'' this will be a differential operator
$X(-\ii \pd,x)$. There are many ways to represent a particular
operator $X$ as a operator function of $p_a$ and $x^a$ which is
related to the \emph{ordering}. The \emph{Weyl ordering} we will
consider in the next subsection, here let us use a different one
in which all operators $p_a$ are on the right to all $x^a$. In
such an ordering prescription one can write down a Fourier
expansion of the operator in the following form
\begin{equation}\label{no}
  X=\frac{1}{(2\pi)^p}\int\dd^p z\, \tilde{X}(z,x) \e^{\ii\hat{p}\cdot z}.
\end{equation}
In this parametrization the product of two operators is given by
an involution product of the symbols:
\begin{equation}\label{star1}
  \widetilde{XY}(z,x)=\tilde{X}*\tilde{Y}(z,x)=\frac{1}{(2\pi)^p}\int
  \dd^p y\, \tilde{X}(y,x)\tilde{Y}(z-y,x+y).
\end{equation}

The trace of an operator can be computed in a standard way, namely
\begin{equation}
  \tr X=\int\dd^p x
  \bra{x}X\ket{x}=\int\dd^px\,\tilde{X}(0,x)=\int\dd^px\dd^pl\,
  X(l,x),
\end{equation}
where in the last part $X(l,x)$ is the normal symbol of which is
obtained by the replacement of operator $\hat{p_a}$ by an ordinary
variable $l_a$ in the definition \eqref{no},
\begin{align}
  X(l,x)&=\frac{1}{(2\pi)^p}\int\dd^pz \, \tilde{X}(z,x)\e^{\ii l\cdot
  z}, \\
  \tilde{X}(z,x)&=\tr\e^{-\ii\hat{p}\cdot z} X.
\end{align}

Now we are ready to write down the whole matrix action
\eqref{mmncs} in terms of the normal symbols. It looks as follows,
\begin{equation}\label{diff-ym}
  S=\int\dd^pl\dd^p x\,\left( -\ft 14\F_{ab}^2+\ft12(\na_a \Phi_I)^2-
  \ft14[\Phi_I,\Phi_J]_*^2\right),
\end{equation}
where
\begin{align}
  \F_{ab}(l,x)&=\pd_a A_b(l,x)-\pd_b A(l,x)-[A_a,A_b]_*(l,x),\\
  \na_a \Phi&=\pd_a \Phi+[A_a,\Phi]_*(l,x), \\
  [A,B]_*(l,x)&=A*B(l,x)-B*A(l,x)
\end{align}
and the star product is defined as in \eqref{star1}.

The model defined by the action \eqref{diff-ym} has the meaning of
Yang--Mills theory with the infinite dimensional gauge group of
diffeomorphism transformations generated by the operators
\begin{equation}
  T_f=\ii f^a(x)\pd_a.
\end{equation}

Because of the noncommutative nature of the products involved in
the action \eqref{diff-ym} the local gauge group is not
commutative. However, if one tries to write down the group of
global gauge symmetry, one finds out that this group is, in fact
nothing else that U(1). Changing only slightly the character of
the solution one can also get a non-Abelian global group. Indeed,
consider the solution as in \eqref{bg} with the exception that the
Hilbert space is not just \eqref{hs-spl}, but is given by the
product of parts $\hh_a$ at some (positive integer) power $n$:
\begin{equation}
  \hh=\left(\otimes_{a=1}^p \hh_a\right)^{\otimes n}.
\end{equation}
Repeating with this solution the same manipulations which lead us
to \eqref{diff-ym} with the only exception that in this case an
arbitrary matrix is represented by a $(n\times n)$-matrix valued
function instead of just ``ordinary'' one, we arrive to the action
similar to \eqref{diff-ym} with the exception that the fields take
their value in the u$(n)$ algebra and the global gauge group is,
respectively, U$(n)$. We hope, that the things will clarify a lot
when the reader will pass the next subsection.

\subsubsection{Ordinary gauge model?}
A question one may ask oneself is if the fluctuations of the
matrix models can be restricted in such a way to get a ``normal''
Yang--Mills theory with a compact Lie group. In the present case
one may restrict the fluctuations around the background \eqref{bg}
to depend on $\hat{x}^a$ operators only. This aim can be achieved
by imposing the following constraints on the matrices $X_\mu$:
\begin{equation}\label{make-comm}
  [x_a, X_b]=\ii \delta_{ab},\qquad [x_a,X_I]=0.
\end{equation}

Let us note that $X_a$ and $x_a$ do not form the Heisenberg
algebra because the commutator between $X_a$ do not necessarily
vanish:
\begin{equation}
  [X_a,X_b]\equiv F_{ab}\neq 0.
\end{equation}

Dynamically, the constraint \eqref{make-comm} can be implemented
through the modification of the matrix action by the addition of
the constraint \eqref{make-comm} with the Lagrange multiplier. The
modified matrix model action reads:
\begin{equation}\label{mm-comm}
  S_c=\tr\left(\ft 14[X_\mu,X_\nu]^2+\rho_{\mu\nu}([x_\mu,X_\nu]-
  \Delta_{\mu\nu})+
  T^2\right),
\end{equation}
where $\rho_{\mu\nu}$ are the Lagrange multipliers,
$x_\mu=(x_a,0)$ and $\Delta_{\mu\nu}$ is equal to $\delta_{ab}$
when $(\mu\nu)=(ab)$ and zero otherwise. The limit $N\to\infty$ of
the matrix model specified by the action \eqref{mm-comm} produces
the Abelian gauge model. Under similar setup one can obtain also
nonabelian gauge models.

\subsection{Nonzero commutator: Noncommutative Yang--Mills model}

In this subsection we consider the matrix action as a perturbation
of the background configuration given by \eqref{heis} and
\eqref{d-ha}. Here we plan to give a more detailed approach also
partly justifying the result of the previous subsection. The
operators $p_a$ generate a $(p+1)/2$-dimensional Heisenberg
algebra. If this algebra is represented irreducibly on the Hilbert
space of the model (which is in fact our choice), then an
arbitrary operator acting on this space can be represented as an
operator function of $p_a$. Let us consider this situation in more
details.

Irreducibility of the representation in particular means that any
operator commuting with all $p_a$ is a $c$-number constant. From
this follows that the operators
\begin{equation}\label{op-P}
  P_a=[p_a,\cdot ],
\end{equation}
which are Hermitian on the space of square trace operators
equipped with the scalar product $(A,B)=\tr A^* B$, are
diagonalizable and have non-degenerate eigenvalues.
\begin{exercise}
  Prove this!
\end{exercise}
By a direct check one can verify that the operator $\e^{\ii k_a
\hat{x}^a}$, where $\hat{x}^a=\theta^{ab}\hat{p}_b$, $\theta\equiv
B^{-1}$ is an eigenvector for $P_a$ with the eigenvalue $k_a$:
\begin{equation}
  P_a\cdot \e^{\ii k\cdot\hat{x}}=[p_a,\e^{\ii k\cdot\hat{x}}]
  =k_a \e^{\ii k\cdot\hat{x}}.
\end{equation}
This set of eigenvectors form an orthogonal basis ($P_a$'s are
Hermitian). One can normalize the eigenvectors to delta function
trace,
\begin{equation}\label{tr-E}
  E_k=c_k\e^{\ii k\cdot\hat{x}},\qquad \tr E^*_{k'}
  E_k=\delta(k'-k).
\end{equation}
The normalizing coefficients $c_k$ can be found from evaluating
explicitly the trace of $\e^{\ii (k-k')\hat{x}}$ in  \eqref{tr-E}
and equating it to the Dirac delta. Let us compute this trace and
find the respective quotients. To do this, consider the basis
where the set of operators $x^\mu$ splits in pairs $p_i$, $q^i$
satisfying the standard commutation relations: $[p_i,q_j]=-\ii
\theta\delta_{ij}$.

As we know from courses of Quantum Mechanics the trace of the
operator
\begin{equation}\label{e.e}
  \e^{-\ii k'\hat{x}}\cdot \e^{\ii k\hat{x}}=\e^{\ii(k-k')\hat{x}}
  \e^{\frac{\ii}{2}k'\times k},
\end{equation}
can be computed in $q$-representation as,
\begin{equation}\label{qs1}
  \tr\e^{\ii(k-k')\hat{x}}
  \e^{\frac{\ii}{2}k'\times k}=\int\dd q \bra{q}
  \e^{-\ii(l'_i-l_i)q^i+(z^{\prime i}-z^i)p_i}\ket{q}=
  1/|c_k|^2\delta(k'-k),
\end{equation}
where $\ket{q}$ is the basis of eigenvectors of $\mathbf{q}^i$,
\begin{equation}\label{q-rep}
  \mathbf{q}^i\ket{q}=q^i\ket{q},\qquad
  \bracket{q'}{q}=\delta(q'-q),
\end{equation}
and $l_i$, $z^i$ ($l_i$, $z^i$) are components of $k_\mu$
($k'_\mu$) in the in the parameterizations: $x^\mu\to p_i,q^i$.
Explicit computation gives,
\begin{equation}\label{qs2}
  1/|c_k|^2=\frac{(2\pi)^{\frac{p}{2}}}{\sqrt{\det\theta}}.
\end{equation}

Now, we have the basis of eigenvectors $E_k$ and can write any
operator $F$ in terms of this basis,
\begin{equation}\label{F-expand}
  \hat{F}=\int\dd k\, \tilde{F}(k)\e^{\ii k\x},
\end{equation}
where the ``coordinate'' $\tilde{F}(k)$ is given by,
\begin{equation}\label{F-inv}
  \tilde{F}(k)=\frac{\sqrt{\det\theta}}{(2\pi)^{\frac{p}{2}}}
  \tr (\e^{-\ii k\x}\cdot \hat{F}).
\end{equation}

Function $\tilde{F}(k)$ can be interpreted as the Fourier
transform of a $L^2$ function $F(x)$,
\begin{equation}\label{Weyl-F}
  F(x)=\int \dd k \tilde{F}(k) \e^{\ii k_\mu x^\mu}=
  \sqrt{\det\theta}\int \frac{\dd k}{(2\pi)^{p/2}}\e^{\ii
  kx}\tr\e^{-\ii k\x}\hat{F}.
\end{equation}
And viceversa, to any $L^2$ function $F(x)$ from one can put into
correspondence an $L^2$ operator $\hat{F}$ by inverse formula,
\begin{equation}\label{Weyl-inv}
  \hat{F}=\int \frac{\dd x}{(2\pi)^{p/2}} \int\frac{\dd k}{(2\pi)^{p/2}}F(x)
  \e^{\ii k(\x-x)}.
\end{equation}

Equations \eqref{Weyl-F} and \eqref{Weyl-inv} providing a
one-to-one correspondence between $L^2$ functions and operators
with finite trace,
\begin{equation}\label{tr-cl}
  \tr \mathbf{F}^\dag\cdot \mathbf{F}<\infty,
\end{equation}
give in fact formula for the Weyl symbols. By introducing
distributions over this space of operators one can extend the
above map to operators with unbounded trace.
\begin{exercise}
 Check that \eqref{Weyl-F} and \eqref{Weyl-inv} lead in terms of
 distributions to the correct Weyl ordering prescription for polynomial functions of
 $p_\mu$.
\end{exercise}

Let us note, that the map \eqref{Weyl-F} and \eqref{Weyl-inv} can
be rewritten in the following form,
\begin{equation}\label{f-delta}
  F(x)=(2\pi)^{p/2}\sqrt{\det\theta}\tr \dop (\x-x)
  \hat{F},\qquad \hat{F}=\int\dd^px\,\dop (\x-x) F(x),
\end{equation}
where we introduced the operator,
\begin{equation}\label{nc-delta}
  \dop (\x-x)=\int\frac{\dd^pk}{(2\pi)^p}\e^{\ii k\cdot(\x-x)}.
\end{equation}
This operator satisfy the following properties,
\begin{subequations}\label{dop-prop}
\begin{align}
  &\int \dd^px\, \dop(\x-x)=\II,\\
  &(2\pi)^{p/2}\sqrt{\det{\theta}}\tr \dop(\x-x)=1,\\
  &(2\pi)^{p/2}\sqrt{\det{\theta}}\tr \dop(\x-x)\dop(\x-y)=
  \delta(x-y),
\end{align}
where in the r.h.s. of last equation is the ordinary delta
function. Also, operators $\dop (\x-x)$ for all $x$ form a
complete set of operators,
\begin{equation}\label{dop-compl}
  [\dop (\x-x),\mathbf{F}]\equiv 0 \Rightarrow F\propto\I.
\end{equation}
The commutation relations of $\x^\mu$ also imply that $\dop(\x-x)$
should satisfy,
\begin{equation}\label{dop-comm}
  [\x^\mu,\dop(\x-x)]=\ii \theta^{\mu\nu}\pd_\nu\dop(\x-x).
\end{equation}
\end{subequations}

In fact one can define alternatively the noncommutative plane
starting from operator $\dop(\x-x)$ satisfying \eqref{dop-prop},
with $\x^\mu$ defined by,
\begin{equation}\label{x-mu}
  \x^\mu=\int\dd^px\, x^\mu \dop(\x-x).
\end{equation}
In this case \eqref{dop-comm} provides that $\x^\mu$ satisfy the
Heisenberg algebra \eqref{heis}, while the property
\eqref{dop-compl} provides that they form a complete set of
operators. Relaxing these properties allows one to introduce a
more general noncommutative spaces.

Let us the operator $\dop (x)$ in the simplest case of
two-dimensional noncommutative plane. The most convenient is to
find its matrix elements $D_{mn}(x)$ in the oscillator basis given
by,
\begin{equation}
  \ket{n}=\frac{(\hat{a}^\dag)^n}{\sqrt{n!}}\ket{0}, \qquad
  \hat{a}\ket{0}=0,
\end{equation}
where the oscillator operators $\hat{a}$ and $\hat{a}^\dag$ are
the noncommutative analogues of the complex coordinates,
\begin{equation}\label{oop}
  \hat{a}=\sqrt{\frac{1}{2\theta}}(\x^1+\ii \x^2),\quad
  \hat{a}^\dag=\sqrt{\frac{1}{2\theta}}(\x^1-\ii \x^2);\quad
  [\hat{a},\hat{a}^\dag]=1.
\end{equation}

Then the matrix elements read
\begin{equation}\label{dop-osc-bas}
  D_{mn}(x)=\bra{m}\dop^{(2)} (\hat{a}-z)\ket{n}=\tr \dop^{(2)}(\hat{a}-z)
  P_{nm},
\end{equation}
where $P_{nm}=\ket{n}\bra{m}$.

As one can see, up to a Hermitian transposition the matrix
elements of $\dop(\x-x)$ correspond to the Weyl symbols of
operators like $\ket{m}\bra{n}$, or so called Wigner functions.
The computation of \eqref{dop-osc-bas} gives,\footnote{For the
details of computation see e.g. \cite{Harvey:2001yn}.}
\begin{equation}\label{D-matr-el}
  D^\theta_{mn}(z,\bar{z})=(-1)^n\left(\frac{2}{\sqrt{\theta}}\right)^{m-n+1}
  \sqrt{\frac{n!}{m!}} \e^{-z\bar{z}/\theta}
  \left(\frac{z^m}{\bar{z}^n}\right)L_n^{m-n}(2z\bar{z}/\theta),
\end{equation}
where $L_n^{m-n}(x)$ are Laguerre polynomials,
\begin{equation}\label{Lag}
  L_n^\alpha(x)=\frac{x^{-\alpha}\e^x}{n!}
  \left(\frac{\dd}{\dd x}\right)^n(\e^{-x}x^{\alpha+n}).
\end{equation}
It is worthwhile to note that in spite of its singular origin the
symbol of the delta operator is a smooth function which is rapidly
vanishing at infinity. The smoothness comes from the fact that the
operator elements are written in an $L^2$ basis. In a non-$L^2$
basis, e.g. in the basis of $x_1$ eigenfunctions $D^{\theta}$
would have more singular form.

The above computations can be generalized to $p$-dimensions.
Written in the complex coordinates $z_i,\bar{z}_i$ corresponding
to oscillator operators \eqref{oop}, which diagonalize the
noncommutativity matrix this looks as follows,
\begin{equation}\label{D-matr-high}
  D_{\vec{m}\vec{n}}=D^{\theta_{(1)}}_{m_1n_1}(z_1,\bar{z}_1)
  D^{\theta_{(2)}}_{m_2n_2}(z_2,\bar{z}_2)
  \dots
  D^{\theta_{(p/2)}}_{m_{p/2}n_{p/2}}(z_{p/2},\bar{z}_{p/2}),
\end{equation}
where,
\begin{equation}\label{zz}
  [z_i,\bar{z}_j]_*=\delta_{ij},\qquad i=1,\dots,p/2.
\end{equation}

Having the above map one can establish the following relations
between operators and their Weyl symbols.
\begin{enumerate}
\item It is not difficult to derive that,
\begin{equation}\label{tr-int}
  (2\pi)^{p/2}\sqrt{\det\theta}\tr\mathbf{F}=\int \dd x\, F(x).
\end{equation}
\item The (noncommutative) product of operators is mapped into the
\emph{star} or \emph{Moyal} product of functions,
\begin{equation}\label{star}
  \mathbf{F}\cdot \mathbf{G}\to F*G(x),
\end{equation}
where $F*G(x)$ is defined as,
\begin{equation}\label{star-def}
  F*G(x)=\left.\e^{-\frac{\ii}{2}\theta^{\mu\nu}\pd_\mu\pd'_\nu}
  F(x)G(x')\right|_{x'=x}.
\end{equation}
In terms of operator $\dop(\x-x)$, this product can be written as
follows,
\begin{equation}\label{star-delta}
  F*G(x)=\int\dd^py\dd^pz\, K(x;y,z)F(y)G(z),
\end{equation}
where,
\begin{multline}\label{K}
  K(x;y,z)=\\
  (2\pi)^{p/2}\sqrt{\det{\theta}}\tr
  \dop(\x-x)\dop(\x-y)\dop(\x-z)=\\
  \e^{\frac{\ii}{2}\pd^y_\mu\theta^{\mu\nu}\pd^z_\nu}
  \delta(y-x)\delta(z-x),
\end{multline}
$\pd^y_\mu$ and $\pd^z_\mu$ are, respectively, $\pd/\pd y^\mu$ and
$\pd/\pd z^\mu$, and in the last line one has ordinary delta
functions.

On the other hand the ordinary product of functions was not found
to have any reasonable meaning in this context.

\item One property of the star product is that in the integrand
one can drop it once because of,
\begin{equation}\label{*-drop}
  \int \dd^px\, F*G(x)=\int \dd^px\,F(x)G(x),
\end{equation}
were in the r.h.s the ordinary product is assumed.

\item Interesting feature of this representation is that partial
derivatives of Weyl symbols correspond to commutators of
respective operators with $\ii \mathbf{p}_\mu$,
\begin{equation}\label{deriv}
  [\ii \mathbf{p}_\mu,\mathbf{F}]\to \ii(p_\mu*F-F*p_\mu)(x)
  =\frac{\pd F(x)}{\pd x^\mu},
\end{equation}
where $p_\mu$ is linear function of $x^\mu$:
$p_\mu=-\theta^{-1}_{\mu\nu}x^\nu$.
\end{enumerate}

This is an important feature of the star algebra of functions
distinguishing it from the ordinary product algebra. In the last
one can not represent the derivative as an \emph{internal
automorphism} while in the star algebra it is possible due to its
nonlocal character. This property is of great importance in the
field theory since, as it will appear later, it is the source of
duality relations in noncommutative gauge models which we turn to
in the next section.

\begin{exercise}
 Derive equations \eqref{tr-int}--\eqref{deriv}.
\end{exercise}

Let us turn back to the matrix model action \eqref{mmncs} and
represent an arbitrary matrix configuration as a perturbation of
the background \eqref{heis}:
\begin{equation}
  X_a=p_a+A_a,\quad X_I=\Phi_I,\qquad a=1,\dots,p+1,\quad
  I=p+2,\dots,D.
\end{equation}

Passing from operators $A_a$ and $\Phi$ to their Weyl symbols
using \eqref{f-delta}, \eqref{star} and \eqref{deriv} one gets
following representation for the matrix action \eqref{mmncs}:
\begin{equation}\label{nc-ym}
  S=\int\dd^p x\,\left( -\ft 14(\F_{ab}-B_{ab})^2+\ft12(\na_a \Phi_I)^2-
  \ft14[\Phi_I,\Phi_J]_*^2\right),
\end{equation}
where,
\begin{equation}\label{f-mn}
  F_{\mu\nu}=\pd_\mu A_\nu-\pd_\nu A_\mu-\ii[A_\mu,A_\nu]_*.
\end{equation}

In the case of the irreducible representation of the algebra
\eqref{d-ha} this describes the U(1) gauge model.

One can consider a $n$-tuple degenerate representation in this
case as well. As in the previous case the index labelling the
representations become an internal symmetry index and the global
gauge group of the model becomes U$(n)$. Indeed, the operator
basis in which one can expand an arbitrary operator now is given
by,
\begin{equation}
  E^{\alpha}_k=\sigma^\alpha\otimes\e^{\ii k\cdot\hat{x}},
\end{equation}
where $\sigma^\alpha$, $\alpha=1,\dots,n^2$ are the adjoint
generators of the u$(n)$ algebra. The can be normalized to
satisfy,
\begin{equation}\label{su-alg}
  [\sigma^\alpha,\sigma^\beta]=\ii
  \epsilon^{\alpha\beta\gamma}\sigma^\gamma, \qquad
  \tr_{su(2)}\sigma^\alpha\sigma^\beta=\delta^{\alpha\beta},
\end{equation}
where $\epsilon^{\alpha\beta\gamma}$ are the structure constants
of the u$(n)$ algebra:
\begin{equation}
 \epsilon^{\alpha\beta\gamma}=-\ii
 \tr_{su(2)}[\sigma^\alpha,\sigma^\beta]\sigma^\gamma,
\end{equation}
which follows from \eqref{su-alg}. Then an operator $\hat{F}$ is
mapped to the following function $F(x)$:
\begin{multline}\label{gWeyl}
  F^\alpha(x)=\sqrt{\det\theta}\int \frac{\dd^p
  k}{(2\pi)^{p/2}}\e^{\ii kx} \tr \left\{(\sigma^\alpha\otimes \e^{\ii
  k\cdot \hat{x}})\cdot \hat{F}\right\}=\\
  (2\pi)^{p/2}\sqrt{\det\theta}
  \tr\left\{(\sigma^\alpha\otimes \dop(\hat{x}-x))\cdot \hat{F}\right\}.
\end{multline}
The equation \eqref{gWeyl} gives the most generic map from the
space of operators to the space of $p$-dimensional u$(n)$-algebra
valued functions.

\begin{exercise}
  Prove that $p$ is always even.
\end{exercise}

Just for the sake of completeness let us give also the formula for
the inverse map,
\begin{equation}\label{gWeyl-inv}
  \hat{F}=\int\dd^px\,(\sigma^\alpha\otimes\dop (\x-x)) F^\alpha(x),
\end{equation}

Applying the map \eqref{gWeyl} and \eqref{gWeyl-inv} to the IKKT
matrix model \eqref{mmncs} or to the BFSS one \eqref{bfss-a}, one
gets, respectively, the $p$ or $p+1$ dimensional noncommutative
u$(n)$ Yang--Mills model.
\begin{exercise}
  Derive the $p$- and $(p+1)$-dimensional noncommutative supersymmetric gauge
  model from the matrix actions \eqref{mmncs} and \eqref{bfss-a},
  using the map \eqref{gWeyl} and its inverse \eqref{gWeyl-inv}.
\end{exercise}

Some comments regarding both gauge models described by the actions
\eqref{diff-ym} and \eqref{f-mn} are in order. In spite of the
fact that both models look very similar to the ``ordinary''
Yang--Mills models, the perturbation theory of this models are
badly defined in the case of noncompact noncommutative spaces. In
the first case the non-renormalizable divergence is due to extra
integrations over $l$ in the ``internal'' space. In the case of
noncommutative gauge model the behavior of the perturbative
expansion is altered by the IR/UV mixing
\cite{Minwalla:1999px,VanRaamsdonk:2000rr}. The supersymmetry or
low dimensionality improves the situation allowing the ``bad''
terms to cancel (see
\cite{Sarkar:2002pb,Bietenholz:2002ch,Slavnov:2003ae,Buric:2003qv}).
On the other hand the compact noncommutative spaces provide both
IR and UV cut off and the field theory on such spaces is finite
\cite{Sheikh-Jabbari:1999iw}. In the case of zero commutator
background the behavior of the perturbative expansion depends on
the eigenvalue distribution. Faster the eigenvalues increase,
better the expansion converge. However there is always the problem
of the zero modes corresponding to the diagonal matrix excitations
(functions of commutative $p_a$'s). There is a hope that
integrating over the remaining modes helps to generate a dynamical
term for the zero modes too. Indeed, for purely bosonic model one
has a repelling potential after the one-loop integration of the
non-diagonal modes. The fermions contribute with the attractive
potential. In the supersymmetric case the repelling bosonic
contribution is cancelled by the attractive fermionic one
 and diagonal modes remain non-dynamical \cite{Makeenko:privat}.

\begin{exercise}
  Consider the Eguchi--Kawai model given by the action \eqref{ek}.
  Write down the equations of motion and find the classical
  solutions analogous to \eqref{heis}. One can have noncommutative
  solutions even for finite $N$. Explain, why? Consider arbitrary
  matrix configuration as a perturbation of the above classical
  backgrounds and find the resulting models. What is the space on
  which this models live? How the same space can be obtained from
  a non-compact matrix model.
\end{exercise}

We considered exclusively the bosonic models. When the
supersymmetric theories are analyzed one has to deal also with the
fermionic part. In the case of compact noncommutative spaces which
correspond to finite size matrices one has a discrete system with
fermions. In the lattice gauge theories with fermions there is a
famous problem related to the fermion \emph{doubling}
\cite{Nielsen:1981hk}. Concerning the theories on the compact
noncommutative spaces it was found that in some cases one can
indeed have fermion doubling \cite{Sochichiu:2000fs}\footnote{For
the case of the unitary Eguchi--Kawai-type model with fermions see
\cite{Kitsunezaki:1997iu}} some other cases were reported to be
doubling free and giving alternative solutions to the long
standing lattice problem \cite{Balachandran:2003ay}.

\section{Matrix models and dualities of noncommutative gauge models}

In the previous section we realized that the matrix model from
different ``points'' of the moduli space of classical solutions
looks like different gauge models. These models can have different
dimensionality or different global gauge symmetry group, but they
all are equivalent to the original IKKT or BFSS matrix model. This
equivalence can be used to pass from some noncommutative model
back to the matrix model and then to a different noncommutative
model and viceversa. Thus, one can find a one-to-one map from one
model to an equivalent one.

In reality one can jump the intermediate step by writing a new
solution direct in the noncommutative gauge model and passing to
Weyl (re)ordered description with respect to the new background.
From the point of view of noncommutative geometry this procedure
is nothing else that the change of the noncommutative variable
taking into account also the ordering. Let us go to the details.
Consider two different background solutions given by
$p^{(i)}_{\mu_{(i)}}$,  where $\mu_{(i)}=1,\dots,p_{(i)}$ and the
index $i=1,2$ labels the backgrounds. Denote the orders of
degeneracy of the backgrounds by $n_{(i)}$. The commutator for
both backgrounds is given by,
\begin{equation}
  [p^{(i)}_{\mu_{(i)}},p^{(i)}_{\nu_{(i)}}]=
  \ii B^{(i)}_{\mu_{(i)}\nu_{(i)}}.
\end{equation}

Applying to a $p_{(1)}$-dimensional u$(n_{(1)}$ algebra valued
field $F^{\alpha_{(1)}}(x_{(1)})$ first the inverse Weyl
transformation \eqref{gWeyl-inv} which maps it in the operator
form and then the direct transformation \eqref{gWeyl} from the
operator form to the second background one gets a
$p_{(2)}$-dimensional u$(n_{(2)}$ algebra valued field
$F^{\alpha_{(2)}}(x_{(2)})$ defined by
\begin{equation}\label{transf}
  F^{\alpha_{(2)}}(x_{(2)})=\int\dd^{p_{(1)}}x_{(1)}
  K_{(2|1)}^{\alpha_{(2)}\alpha_{(1)}}(x_{(2)}|x_{(1)})F^{\alpha_{(1)}}(x_{(1)}),
\end{equation}
where the kernel
$K_{(2|1)}^{\alpha_{(2)}\alpha_{(1)}}(x_{(2)}|x_{(1)})$ is given
by,
\begin{multline}\label{kern1}
  K_{(1|2)}^{\alpha_{(2)}\alpha_{(1)}}(x_{(2)},x_{(1)})=
  (2\pi)^{p_{(2)}/2}\sqrt{\det\theta_{(2)}}\times \\
  \tr\left\{(\sigma_{(2)}^{\alpha_{(2)}}\otimes \dop(\hat{x}_{(2)}-x_{(2)}))
  \cdot(\sigma_{(1)}^{\alpha_{(1)}}
   \otimes\dop (\x_{(1)}-x_{(1)}))
   \right\},
\end{multline}
where $x_{(i)}$ and $\sigma_{(i)}^{\alpha_{(i)}}$ are the
coordinate and algebra generators corresponding to the background
$p^{(i)}_{\mu_{(i)}}$.

The equation \eqref{kern1} still appeals to the background
independent operator form by using the $\dop$-operators and trace.
This can be eliminated in the following way. Consider the
functions
$x_{(2)}^{\mu_{(2)}}(x_{(1)}^{\mu_{(1)}},\sigma_{(1)}^{\alpha_{(1)}})=
x_{(2)}^{\mu_{(2)};\alpha_{(1)}}(x_{(1)}^{\mu_{(1)}})\sigma_{(1)}^{\alpha_{(1)}}$
and
$\sigma_{(2)}^{\alpha_{(2)}}(x_{(1)}^{\mu_{(1)}},\sigma_{(1)}^{\alpha_{(1)}})=
\sigma_{(2)}^{\alpha_{(2)};\alpha_{(1)}}(x_{(1)}^{\mu_{(1)}})\sigma_{(1)}^{\alpha_{(1)}}$
which are the symbols of the second background
$\x_{(2)}^{\mu_{(2)}}$ which are Weyl-ordered with respect to the
first background. Namely, they are the solution to the equation,
\begin{equation}
  x_{(2)}^{\mu_{(2)}}*_{(1)}x_{(2)}^{\nu_{(2)}}
  -
  x_{(2)}^{\nu_{(2)}}*_{(1)}x_{(2)}^{\mu_{(2)}}
  =
  \theta^{(2)}_{\mu_{(2)}\nu_{(2)}},
\end{equation}
and for $\sigma_{(2)}$
\begin{equation}
   \sigma_{(2)}^{\alpha_{(2)}}*_{(1)}\sigma_{(2)}^{\beta_{(2)}}
  -
  \sigma_{(2)}^{\alpha_{(2)}}*_{(1)}\sigma_{(2)}^{\beta_{(2)}}
  =
  \ii\epsilon^{\alpha_{(2)}\beta_{(2)}\gamma_{(2)}}\sigma_{(2)}^{\gamma_{(2)}}
\end{equation}
where $*_{(1)}$ includes both the noncommutative with
$\theta_{(1)}$ and the u$(n_{(1)})$ matrix products and we did not
write explicitly the arguments
$(x_{(1)}^{\mu_{(1)}},\sigma_{(1)}^{\alpha_{(1)}})$ and
u$(n_{(1)})$ matrix indices of $x_{(2)}$ and $\sigma_{(2)}$. Then,
the kernel \eqref{kern1} can be rewritten in the $x_{(1)}$
background as follows,
\begin{multline}\label{kern2}
  K_{(1|2)}^{\alpha_{(2)}\alpha_{(1)}}(x_{(2)},x_{(1)})=\\
  \sqrt{\frac{\det 2\pi\theta_{(2)}}{\det 2\pi\theta_{(1)}}}
  d_{(1)}^{\alpha_{(1)}\beta_{(1)}\gamma_{(1)}}\left(\sigma_{(2)}^{\alpha_{(2)};\beta_{(1)}}
  *_{(1)}\delta_{*_{(1)}}^{\gamma_{(1)}}(x_{(2)}(x_{(1)})-x_{(2)})
  \right),
\end{multline}
where
$d_{(1)}^{\alpha\beta\gamma}=\tr_{(1)}\sigma_{(1)}^\alpha\sigma_{(1)}^\beta
\sigma_{(1)}^\gamma$ and
\begin{equation}
   \delta_{*_{(1)}}^{\gamma_{(1)}}(x_{(2)}(x_{(1)})-x_{(2)})=
   \int\frac{\dd^{p_{(2)}}l}{(2\pi)^{p_{(2)}}}
   \tr_{(1)}\sigma_{(1)}^{\gamma_{(1)}}\e_{*_{(1)}}^{\ii l\cdot
   (x_{(2)}(x_{(1)})-x_{(2)})},
\end{equation}
$\e_*^{f(x)}$ is the star exponent computed with the
noncommutative structure corresponding to $*$.

General expression for the basis transform \eqref{transf} with the
kernel \eqref{kern1} or \eqref{kern2} looks rather complicate
almost impossible to deal with. Therefore it is useful to consider
some particular examples which we take from \cite{Kiritsis:1997hj}
which show that in fact the objects are still treatable.

\subsection{Example 1: The U(1) $\longrightarrow$ U($n$) map}
Let us present the explicit construction for the map from U(1) to
U(2) gauge model in the case of two-dimensional non-commutative
space. The map we are going to discuss can be straightforwardly
generalised to the case of arbitrary even dimensions as well as to
the case of arbitrary U(n) group.

The two-dimensional non-commutative coordinates are,
\begin{equation}\label{2d}
  [x^1,x^2]=\ii \theta.
\end{equation}

As we already discussed, non-commutative analog of complex
coordinates is given by oscillator rising and lowering operators,
\begin{gather}\label{aabar}
  a=\sqrt{\frac{1}{2\theta}}(x^1+\ii x^2),\qquad
  \bar{a}=\sqrt{\frac{1}{2\theta}}(x^1-\ii x^2)\\
  a\ket{n}=\sqrt{n}\ket{n-1},\qquad \bar{a}\ket{n}=\sqrt{n+1}\ket{n+1},
\end{gather}
where $\ket{n}$ is the oscillator basis formed by eigenvectors of
$N=\bar{a}a$,
\begin{equation}\label{n}
  N\ket{n}=n\ket{n}.
\end{equation}
The gauge symmetry in this background is non-commutative U(1).

We will now construct the non-commutative U(2) gauge model. For
this, consider the U(2) basis which is given by following vectors,
\begin{gather}\label{hv}
  \ket{n',a}=\ket{n'}\otimes e_a, \qquad a=0,1\\ e_0= \begin{pmatrix}
  1\\0 \end{pmatrix},\qquad e_1= \begin{pmatrix} 0\\1 \end{pmatrix},
\end{gather}
where $\{\ket{n'}\}$ is the oscillator basis and $\{e_a\}$ is the
``isotopic'' space basis.

The one-to-one correspondence between U(1) and U(2) bases can be
established in the following way \cite{Nair:2001rt},
\begin{equation}\label{2->1}
  \ket{n'}\otimes e_a\sim \ket{n}=\ket{2n'+a},
\end{equation}
where $\ket{n}$ is a basis element of the U(1)-Hilbert space and
$\ket{n'}\otimes e_a$ is a basis element of the Hilbert space of
U(2)-theory. (Note, that they are two bases of the same Hilbert
space.)

Let us note that the identification \eqref{2->1} is not unique.
For example, one can put an arbitrary unitary matrix in front of
$\ket{n}$ in the r.h.s. of \eqref{2->1}. This in fact describes
all possible identifications and respectively maps from U(1) to
U(2) model.

Under this map, the U(2) valued functions can be represented as
scalar functions in U(1) theory. For example, constant U(2)
matrices are mapped to particular functions in U(1) space. To find
these functions, it suffices to find the map of the basis of the
u$(2)$ algebra given by Pauli matrices $\sigma_\alpha$,
$\alpha=0,1,2,3$.

In the U$(1)$ basis Pauli matrices look as follows,
\begin{subequations}\label{sigma}
 \begin{align} &\sigma_0=\sum_{n=0}^{\infty}
  \bigl(\ket{2n}\bra{2n}+\ket{2n+1}\bra{2n+1}\bigr)\equiv\II,\\
  &\sigma_1=\sum_{n=0}^{\infty} \bigl(\ket{2n}\bra{2n+1}+
  \ket{2n+1}\bra{2n}\bigr),\\ &\sigma_2=-\ii\sum_{n=0}^{\infty}
  \bigl(\ket{2n}\bra{2n+1}- \ket{2n+1}\bra{2n}\bigr),\\
  &\sigma_3=\sum_{n=0}^{\infty} \bigl(\ket{2n}\bra{2n}-
  \ket{2n+1}\bra{2n+1}\bigr), \end{align}
\end{subequations}
while the ``complex'' coordinates $a'$ and $\bar{a}'$ of the U(2)
invariant space are given by the following,
\begin{subequations}\label{aprim}
\begin{align}
  &a'=\sum_{n=0}^{\infty}\sqrt{n}\bigl(\ket{2n-2}\bra{2n}+
  \ket{2n-1}\bra{2n+1}\bigr),\\
  &\bar{a}'=\sum_{n=0}^{\infty}\sqrt{n+1}\bigl(\ket{2n+2}\bra{2n}
  +\ket{2n+3}\bra{2n+1}\bigr).
\end{align}
\end{subequations}

One can see that when trying to find the Weyl symbols for
operators given by \eqref{sigma}, \eqref{aprim}, one faces the
problem that the integrals defining the Weyl symbols diverge.
This happens because the respective functions (operators) do not
belong to the non-commutative analog of $L^2$ space (are not
square-trace).

Let us give an alternative way to compute the functions
corresponding to operators \eqref{sigma} and \eqref{aprim}. To do
this let us observe that operators
\begin{equation}\label{Pi+}
  \Pi_+=\sum_{n=0}^{\infty}\ket{2n}\bra{2n},
\end{equation}
and
\begin{equation}\label{Pi-}
  \Pi_-=\sum_{n=0}^{\infty}\ket{2n+1}\bra{2n+1},
\end{equation}
can be expressed as\footnote{Weyl symbols of $a$ and $\bar{a}$ are
denoted, respectively, as $z$ and $\bar{z}$. The same rule applies
also to primed variables.}
\begin{equation}\label{P+aa}
  \Pi_+=\frac{1}{2}\sum_{n=0}^{\infty}\left(1+
  \sin\pi\left(n+\frac{1}{2}\right)\right)\ket{n}\bra{n}\to
  \frac{1}{2}\left(1+\sin_*\pi\left(\bar{z}*z+\frac{1}{2}\right)\right),
\end{equation}
and,
\begin{equation}\label{P-aa}
  \Pi_-=\I-\Pi_+=
  \frac{1}{2}\left(1-\sin_*\pi\left(\bar{z}*z+\frac{1}{2}\right)\right)=
  \frac{1}{2}\left(1-\sin_*\pi|z|^2\right),
\end{equation}
where $\sin_*$ is the ``star'' sin function defined by the star
Taylor series,
\begin{equation}\label{star-sin}
  \sin_* f=f-\frac{1}{3!}f*f*f+\frac{1}{5!}f*f*f*f*f-\cdots,
\end{equation}
with the star product defined in variables $z,\bar{z}$ as follows,
\begin{equation}\label{aa*}
  f*g(\bar{a},a)=\e^{\pd\bar{\pd}'-\bar{\pd}\pd'}f(\bar{z},z)
  g(\bar{z}',z')|_{z'=z},
\end{equation}
where $\pd=\pd/\pd z$, $\bar{\pd}=\pd/\pd\bar{z}$ and analogously
for primed $z'$ and $\bar{z}'$. For convenience we denoted Weyl
symbols of $a$ and $\bar{a}$ as $z$ and $\bar{z}$.

The easiest way to compute \eqref{P+aa} and \eqref{P-aa} is to
find the Weyl symbol of the operator,
\begin{equation}\label{I}
  I^{\pm}_k=\frac{1\pm\sin\left(\bar{a}a+
  \frac{1}{2}\right)}{(\bar{a}a+\gamma)^k},
\end{equation}
were $\gamma$ is some constant, mainly $\pm 1/2$.

For sufficiently large $k$, the operator $I^\pm_k$ becomes square
trace for which the formula \eqref{gWeyl} defining the Weyl map is
applicable. The result can be analytically continued for smaller
values of $k$, using the following recurrence relation,
\begin{equation}\label{rec}
  I^{\pm}_{k-m}(\bar{z},z)=
  \underbrace{\left(|z|^2+\gamma-\frac{1}{2}\right)*\dots*
  \left(|z|^2+\gamma-\frac{1}{2}\right)}_{m\text{ times}}
  *I^{\pm}_{k}(\bar{z},z).
\end{equation}
The last equation requires computation of only finite number of
derivatives of $I^\pm_{k}(\bar{z},z)$ arising from the star
product with polynomials in $\bar{z},z$.

\begin{exercise}
  Compute the Weyl symbol for the operator \eqref{I}.
\end{exercise}

\subsection{Example 2: Map between different dimensions}

Consider the situation when the dimension is changed. This topic
was considered in \cite{Sochichiu:2000bg,Sochichiu:2000kz}.

Consider the Hilbert space $\hh$ corresponding to the
representation of the two-dimensional non-commutative algebra
\eqref{2d}, and $\hh\otimes\hh$ (which is in fact isomorphic to
$\hh$) which corresponds to the four-dimensional non-commutative
algebra generated by
\begin{equation}\label{4d}
  [x^1,x^2]=\ii\theta_{(1)},\qquad [x^3,x^4]=\ii\theta_{(2)}.
\end{equation}
In the last case non-commutative complex coordinates correspond to
two sets of oscillator operators,  $a_1$, $a_2$ and $\bar{a}_1$,
$\bar{a}_2$, where,
\begin{subequations}\label{4d-osc}
\begin{align}
  &a_1=\sqrt{\frac{1}{2\theta_{(1)}}}(x^1+\ii x^2),\qquad
  &\bar{a}_1=\sqrt{\frac{1}{2\theta_{(1)}}}(x^1-\ii x^2)\\
  &a_1\ket{n_1}=\sqrt{n_1}\ket{n_1-1},\qquad
  &\bar{a}_1\ket{n_1}=\sqrt{n_1+1}\ket{n_1+1},\\
  &a_2=\sqrt{\frac{1}{2\theta_{(2)}}}(x^3+\ii x^4),\qquad
  &\bar{a}_2=\sqrt{\frac{1}{2\theta_{(2)}}}(x^3-\ii x^4)\\
  &a\ket{n}_2=\sqrt{n_2}\ket{n_2-1},\qquad
  &\bar{a}_2\ket{n_2}=\sqrt{n_2+1}\ket{n_2+1},
\end{align}
\end{subequations}
and the basis elements of the ``four-dimensional'' Hilbert space
$\hh\otimes\hh$ are $\ket{n_1,n_2}=\ket{n_1}\otimes\ket{n_2}$.

The isomorphic map $\sigma:\hh\otimes\hh\to\hh$ is given by
assigning a unique number $n$ to each element $\ket{n_1,n_2}$ and
putting it into correspondence to $\ket{n}\in\hh$. So, the
problems is reduced to the construction of an isomorphic map from
one-dimensional lattice of e.g. nonnegative integers into the
two-dimensional quarter-infinite lattice. This can be done by
consecutive enumeration of the two-dimensional lattice nodes
starting from the angle $(00)$. The details of the construction
can be found in Refs. \cite{Sochichiu:2000bg,Sochichiu:2000kz}.

As we discussed earlier, this map induces an isomorphic map of
gauge and scalar fields from two to four dimensional
non-commutative spaces.

This can be easily generalized to the case with arbitrary number
of factors $\hh\otimes\dots\otimes\hh$ corresponding to $p/2$
``two-dimensional'' non-commutative spaces. In this way, one
obtains the isomorphism $\sigma$ which relates two-dimensional
non-commutative function algebra with a $p$-dimensional one, for
$p$ even.

\section{Example 3: Change of $\theta$}

So far, we have considered maps which relate algebras of
non-commutative functions in different dimensions or at least
taking values in different Lie algebras.  Due to the fact that
they change considerably the geometry, these maps could not be
deformed smoothly into the identity map. In this section we
consider a more restricted class of maps which do not change
either dimensionality or the gauge group but only the
non-commutativity parameter. Obviously, this can be smoothly
deformed into identity map, therefore one may consider
infinitesimal transformations.

The new non-commutativity parameter is given by the solution to
the equations of motion. In this framework, the map is given by
the change of the background solution $p_\mu$ by an infinitesimal
amount: $p_\mu+\delta p_\mu$. Then, a solution with the constant
field strength $F^{(\delta p)}_{\mu\nu}$ will change the
non-commutativity parameter as follows,
\begin{equation}\label{delta-theta}
  \theta^{\mu\nu}+\delta
  \theta^{\mu\nu}\equiv(\theta^{-1}_{\mu\nu}+\delta\theta^{-1}_{\mu\nu})^{-1}=
  (\theta^{-1}_{\mu\nu}+F_{\mu\nu})^{-1}.
\end{equation}
Note, that the above equation does not require $\delta\theta$ to
be infinitesimal.

Since we are considering solutions to the gauge field equations of
motion $A_\mu=\delta p_\mu$ one should fix the gauge for it. A
convenient choice would be e.g. the Lorentz gauge, $\pd_\mu \delta
p_\mu=0$. Then, the solution with
\begin{equation}\label{delta-p}
  A^{(\delta p)}_\mu\equiv\delta p_\mu =(1/2)
  \epsilon_{\mu\nu}\theta^{\nu\alpha}p_\alpha
\end{equation}
with antisymmetric $\epsilon_{\mu\nu}$ has the constant field
strength
\begin{equation}\label{F-delta}
 F^{(\delta p)}_{\mu\nu}\equiv
 \delta\theta^{-1}_{\mu\nu}=\epsilon_{\mu\nu}+(1/4)\epsilon_{\mu\alpha}
 \theta^{\alpha\beta}\epsilon_{\beta\nu}=\epsilon_{\mu\nu}+O(\epsilon^2).
\end{equation}
This corresponds to the following variation of the
non-commutativity parameter,
\begin{equation}\label{theta-new}
  \delta \theta^{\mu\nu}=-\theta^{\mu\alpha}\epsilon_{\alpha\beta}
  \theta^{\beta\nu}-\frac{1}{4}\theta^{\mu\alpha}\epsilon_{\alpha\gamma}
  \theta^{\gamma\rho}\epsilon_{\rho\beta}\theta^{\beta\nu}=
  -\theta^{\mu\alpha}\delta\theta^{-1}_{\alpha\beta}
  \theta^{\beta\nu}+O(\epsilon^2).
\end{equation}
Let us note that such kind of infinitesimal transformations were
considered in a slightly different context in
\cite{Ishikawa:2001mq}.

Let us find how non-commutative functions are changed with respect
to this transformation. In order to do this, let us consider how
the Weyl symbol \eqref{gWeyl} transforms under the variation of
background \eqref{delta-p}.  For an arbitrary operator $\phi$
after short calculation we have,
\begin{equation}\label{delta-Phi}
  \delta \phi (x)=\frac{1}{4} \delta\theta^{\alpha\beta}(\pd_\alpha
  \phi*p_\beta(x)+p_\beta*\pd_\alpha \phi (x)).
\end{equation}
In obtaining this equation we had to take into consideration the
variation of $p_\mu$ as well as of the factor $\sqrt{\det\theta}$
in the definition of the Weyl symbol \eqref{gWeyl}.

By the construction, this variation satisfies the  ``star-Leibnitz
rule'',
\begin{equation}\label{prop1}
  \delta (\phi*\chi)(x)=\delta
  \phi*\chi(x)+\phi*\delta\chi(x)+\phi(\delta*)\chi(x),
\end{equation}
where $\delta\phi(x)$ and $\delta\chi(x)$ are defined according to
\eqref{delta-Phi} and variation of the star-product is given by,
\begin{equation}\label{delta-*}
  \phi(\delta*)\chi(x)=\frac{1}{2}\delta\theta^{\alpha\beta}
  \pd_\alpha\phi*\pd_\beta\chi(x).
\end{equation}
The property \eqref{prop1} implies that $\delta$ provides an
homomorphism (which is in fact an isomorphism) of star algebras of
functions.

The above transformation \eqref{delta-Phi} do not apply, however,
to the gauge field $A_\mu(x)$ and gauge field strength
$F_{\mu\nu}(x)$. This is the case because the respective fields do
not correspond to invariant operators. Indeed, according to the
definition $A_\mu=X_\mu-p_\mu$, where $X_\mu$ is corresponds to
such an operator. Therefore, the gauge field $A_\mu(x)$ transforms
in a nonhomogeneous way,\footnote{In fact the same happens in the
map between different dimensions.}
\begin{equation}\label{delta-A}
  \delta A_\mu(x)=\frac{1}{4}\delta\theta^{\alpha\beta} (\pd_\alpha
  A_\mu*p_\beta+p_\beta*\pd_\alpha A_\mu)+
  \frac{1}{2}\theta_{\mu\alpha}\delta\theta^{\alpha\beta}p_\beta.
\end{equation}

The transformation law for $F_{\mu\nu}(x)$ can be computed using
its definition \eqref{f-mn} and the ``star-Leibnitz rule''
\eqref{prop1} as well as the fact that it is the Weyl symbol of
the operator,
\begin{equation}\label{xx->F}
  F_{\mu\nu}=\ii[X_\mu,X_\nu]-\theta_{\mu\nu}.
\end{equation}
Of course, both approaches give the same result,
\begin{equation}\label{delta-F}
  \delta F_{\mu\nu}(x)=\frac{1}{4}\delta\theta^{\alpha\beta}
  (\pd_\alpha F_{\mu\nu}*p_\beta+p_\beta*\pd_\alpha
  F_{\mu\nu})(x)-\delta\theta^{-1}_{\mu\nu}.
\end{equation}

The infinitesimal map described above has the following
properties:
\begin{enumerate}
  \item[$i$).] It maps gauge equivalent configurations to gauge
equivalent ones, therefore it satisfies the Seiberg--Witten
equation,
\begin{equation}\label{SWE}
  U^{-1}*A*U+U^{-1}*dU\to U^{\prime-1}*'A'*'U'+U^{\prime-1}*'d'U'.
\end{equation}
  \item[$ii$).] It is linear in the fields.  \item[$iii$).] Any
  background independent functional is invariant under  this
  transformation. In particular, any gauge invariant functional whose
  dependence on gauge fields enters through the combination
  $X_{\mu\nu}(x)=F_{\mu\nu}+\theta^{-1}_{\mu\nu}$ is invariant with
  respect to \eqref{delta-Phi}--\eqref{delta-F}. This is also the
  symmetry of the action provided that the gauge coupling transforms
  accordingly.  \item[$iv$).] Formally, the transformation
  \eqref{delta-Phi} can be represented in the form,
\begin{equation}\label{xf}
  \delta \phi(x)=\delta x^\alpha \pd_\alpha \phi (x)=\phi (x+\delta
  x)-\phi(x),
\end{equation}
where $\delta x^\alpha=-\theta^{\alpha\beta}\delta p_\beta$ and no
star product is assumed. This looks very similar to the coordinate
transformations.
\end{enumerate}

The map we just constructed looks very similar to the famous
Seiberg--Witten map, which is given by the following variation of
the background $p_\mu$ \cite{Seiberg:1999vs},
\begin{equation}\label{d-pSW}
  \delta_{\mathrm{SW}}p_\mu=-\frac{1}{2}\epsilon_{\mu\nu}
  \theta^{\nu\alpha}A_\alpha.
\end{equation}

In \eqref{delta-p} we have chosen $\delta p_\mu$ independent of
gauge field background. (In fact the gauge field background was
switched-on later, after the transformation.) An alternative way
would be to have nontrivial field $A_\mu(x)$ from the very
beginning and to chose $\delta p_\mu$ to be of the Seiberg--Witten
form. Then, the transformation laws corresponding to such a
transformation of the background coincide exactly with the
standard SW map. This appears possible because the function
$p_\mu=-\theta^{-1}_{\mu\nu}x^\nu$ has the same gauge
transformation properties as $-A_\mu(x)$,
\begin{equation}\label{upu}
  p_\mu\to U^{-1}*p_\mu*U(x)-U^{-1}*\pd_\mu U(x).
\end{equation}

\section{Discussion and outlook}

This lecture notes were designed as a very basic and very
subjective introduction to the field. Many important things were
not reflected and even not mentioned here. Among these, very few
was said about the brane dynamics and interpretation which was the
main motivation for the development of the matrix models, while
the literature on this topic is enormously vast. For this we refer
the reader to other reviews and lecture notes mentioned in the
introduction (as well as to the references one can find inside
these papers).

Recently, the role of the matrix models in the context of AdS/CFT
correspondence became more clear. Some new matrix models arise in
the description of the anomalous dimensions of composite
super-Yang--Mills operators (see e.g
\cite{Agarwal:2004cb,Bellucci:2004fh}.

Another recent progress even not mentioned here but which is
related to matrix models is their use for the computation of the
superpotential of $\mathcal{N}=1$ supersymmetric gauge theories
\cite{Dijkgraaf:2002dh,Dijkgraaf:2002vw,Dijkgraaf:2002fc}.

\textbf{Acknowledgements.} This lecture notes reflects the
experience I gained due to the communication with many persons. My
thanks are directed most of all to my collaborators and colleagues
from Bogoliubov lab in Dubna, Physics Dept of University of Crete,
Laboratori Nazionali di Frascati. The complete list of persons is 
too long.

I am grateful to the friends who helped me with various problems a
while this text was being written: Giorgio Pagnini, Carlo Cavallo,
Valeria and Claudio Minardi.

This work was supported by the INTAS-00-00262 grant and the Alexander
von Humboldt research fellowship.
% ------------------------------------------------------------------
%
%
%%%%%%%%%%%%%%%%%%%%%%%%%%%%%%%%%%%%%%%%%%%%%%%%%%%%%%%%%%%%%%%%%%%%%
%
% Bibtex users please use
\bibliographystyle{hplain}
\bibliography{lect}

\begin{thebibliography}{10}

\bibitem{Agarwal:2004cb}
Abhishek Agarwal and Sarada.~G. Rajeev.
\newblock The dilatation operator of n = 4 sym and classical limits of spin
  chains and matrix models.
\newblock {\em Mod. Phys. Lett.}, A19:2549, 2004, hep-th/0405116.

\bibitem{Aharony:1999ti}
Ofer Aharony, Steven~S. Gubser, Juan~M. Maldacena, Hirosi Ooguri, and Yaron Oz.
\newblock Large n field theories, string theory and gravity.
\newblock {\em Phys. Rept.}, 323:183--386, 2000, hep-th/9905111.

\bibitem{Balachandran:2003ay}
A.~P. Balachandran and Giorgio Immirzi.
\newblock The fuzzy ginsparg-wilson algebra: A solution of the fermion doubling
  problem.
\newblock {\em Phys. Rev.}, D68:065023, 2003, hep-th/0301242.

\bibitem{Banks:1996vh}
Tom Banks, W.~Fischler, S.~H. Shenker, and Leonard Susskind.
\newblock M theory as a matrix model: A conjecture.
\newblock {\em Phys. Rev.}, D55:5112--5128, 1997, hep-th/9610043.

\bibitem{Bellucci:2004fh}
Stefano Bellucci and Corneliu Sochichiu.
\newblock On matrix models for anomalous dimensions of super yang- mills
  theory.
\newblock 2004, hep-th/0410010.

\bibitem{Berenstein:2002zw}
David Berenstein, Edi Gava, Juan~M. Maldacena, K.~S. Narain, and Horatiu
  Nastase.
\newblock Open strings on plane waves and their yang-mills duals.
\newblock 2002, hep-th/0203249.

\bibitem{Berenstein:2002jq}
David Berenstein, Juan~M. Maldacena, and Horatiu Nastase.
\newblock Strings in flat space and pp waves from n = 4 super yang mills.
\newblock {\em JHEP}, 04:013, 2002, hep-th/0202021.

\bibitem{Berenstein:2002sa}
David Berenstein and Horatiu Nastase.
\newblock On lightcone string field theory from super yang-mills and
  holography.
\newblock 2002, hep-th/0205048.

\bibitem{Bietenholz:2002ch}
W.~Bietenholz, F.~Hofheinz, and J.~Nishimura.
\newblock The renormalizability of 2d yang-mills theory on a non- commutative
  geometry.
\newblock {\em JHEP}, 09:009, 2002, hep-th/0203151.

\bibitem{Brody:1981cx}
T.~A. Brody et~al.
\newblock Random matrix physics: Spectrum and strength fluctuations.
\newblock {\em Rev. Mod. Phys.}, 53:385--479, 1981.

\bibitem{Buric:2003qv}
Maja Buric and Voja Radovanovic.
\newblock On renormalizability of the quantum electrodynamics on noncommutative
  space.
\newblock 2003, hep-th/0305236.

\bibitem{Cheung:1998nr}
Yeuk-Kwan~E. Cheung and Morten Krogh.
\newblock Noncommutative geometry from 0-branes in a background b- field.
\newblock {\em Nucl. Phys.}, B528:185--196, 1998, hep-th/9803031.

\bibitem{Chu:1998qz}
Chong-Sun Chu and Pei-Ming Ho.
\newblock Noncommutative open string and d-brane.
\newblock {\em Nucl. Phys.}, B550:151--168, 1999, hep-th/9812219.

\bibitem{Chu:1999gi}
Chong-Sun Chu and Pei-Ming Ho.
\newblock Constrained quantization of open string in background b field and
  noncommutative d-brane.
\newblock {\em Nucl. Phys.}, B568:447--456, 2000, hep-th/9906192.

\bibitem{Dijkgraaf:2002fc}
Robbert Dijkgraaf and Cumrun Vafa.
\newblock Matrix models, topological strings, and supersymmetric gauge
  theories.
\newblock {\em Nucl. Phys.}, B644:3--20, 2002, hep-th/0206255.

\bibitem{Dijkgraaf:2002vw}
Robbert Dijkgraaf and Cumrun Vafa.
\newblock On geometry and matrix models.
\newblock {\em Nucl. Phys.}, B644:21--39, 2002, hep-th/0207106.

\bibitem{Dijkgraaf:2002dh}
Robbert Dijkgraaf and Cumrun Vafa.
\newblock A perturbative window into non-perturbative physics.
\newblock 2002, hep-th/0208048.

\bibitem{Eguchi:1982nm}
Tohru Eguchi and Hikaru Kawai.
\newblock Reduction of dynamical degrees of freedom in the large {N} gauge
  theory.
\newblock {\em Phys. Rev. Lett.}, 48:1063, 1982.

\bibitem{Guhr:1997ve}
Thomas Guhr, Axel Muller-Groeling, and Hans~A. Weidenmuller.
\newblock Random matrix theories in quantum physics: Common concepts.
\newblock {\em Phys. Rept.}, 299:189--425, 1998, cond-mat/9707301.

\bibitem{Harvey:2001yn}
Jeffrey~A. Harvey.
\newblock Komaba lectures on noncommutative solitons and d-branes.
\newblock 2001, hep-th/0102076.

\bibitem{Ishibashi:1996xs}
N.~Ishibashi, H.~Kawai, Y.~Kitazawa, and A.~Tsuchiya.
\newblock A large-n reduced model as superstring.
\newblock {\em Nucl. Phys.}, B498:467--491, 1997, hep-th/9612115.

\bibitem{Ishikawa:2001mq}
Tomomi Ishikawa, Shin-Ichiro Kuroki, and Akifumi Sako.
\newblock Noncommutative cohomological field theory and gms soliton.
\newblock {\em J. Math. Phys.}, 43:872--896, 2002, hep-th/0107033.

\bibitem{Kiritsis:1997hj}
Elias Kiritsis.
\newblock Introduction to superstring theory.
\newblock 1997, hep-th/9709062.

\bibitem{Kiritsis:2002py}
Elias Kiritsis and Corneliu Sochichiu.
\newblock Duality in non-commutative gauge theories as a non- perturbative
  seiberg-witten map.
\newblock 2002, hep-th/0202065.

\bibitem{Kitsunezaki:1997iu}
Naofumi Kitsunezaki and Jun Nishimura.
\newblock Unitary iib matrix model and the dynamical generation of the space
  time.
\newblock {\em Nucl. Phys.}, B526:351--377, 1998, hep-th/9707162.

\bibitem{Makeenko:privat}
Youri Makeenko.
\newblock Privat communication.

\bibitem{Minwalla:1999px}
Shiraz Minwalla, Mark Van~Raamsdonk, and Nathan Seiberg.
\newblock Noncommutative perturbative dynamics.
\newblock {\em JHEP}, 02:020, 2000, hep-th/9912072.

\bibitem{Morozov:2005mz}
A.~Morozov.
\newblock Challenges of matrix models.
\newblock 2005, hep-th/0502010.

\bibitem{Nair:2001rt}
V.~P. Nair and A.~P. Polychronakos.
\newblock On level quantization for the noncommutative chern-simons theory.
\newblock {\em Phys. Rev. Lett.}, 87:030403, 2001, hep-th/0102181.

\bibitem{Nielsen:1981hk}
Holger~Bech Nielsen and M.~Ninomiya.
\newblock No go theorem for regularizing chiral fermions.
\newblock {\em Phys. Lett.}, B105:219, 1981.

\bibitem{Osborn:1998qb}
J.~C. Osborn, D.~Toublan, and J.~J.~M. Verbaarschot.
\newblock From chiral random matrix theory to chiral perturbation theory.
\newblock {\em Nucl. Phys.}, B540:317--344, 1999, hep-th/9806110.

\bibitem{Polchinski:book}
J.~Polchinski.
\newblock {\em String theory}.
\newblock Cambridge, 1998.

\bibitem{Polchinski:1995mt}
Joseph Polchinski.
\newblock Dirichlet-branes and ramond-ramond charges.
\newblock {\em Phys. Rev. Lett.}, 75:4724--4727, 1995, hep-th/9510017.

\bibitem{Sarkar:2002pb}
Swarnendu Sarkar.
\newblock On the uv renormalizability of noncommutative field theories.
\newblock {\em JHEP}, 06:003, 2002, hep-th/0202171.

\bibitem{Seiberg:1999vs}
Nathan Seiberg and Edward Witten.
\newblock String theory and noncommutative geometry.
\newblock {\em JHEP}, 09:032, 1999, hep-th/9908142.

\bibitem{Sheikh-Jabbari:1999iw}
M.~M. Sheikh-Jabbari.
\newblock Renormalizability of the supersymmetric yang-mills theories on the
  noncommutative torus.
\newblock {\em JHEP}, 06:015, 1999, hep-th/9903107.

\bibitem{Shuryak:1992pi}
Edward~V. Shuryak and J.~J.~M. Verbaarschot.
\newblock Random matrix theory and spectral sum rules for the dirac operator in
  qcd.
\newblock {\em Nucl. Phys.}, A560:306--320, 1993, hep-th/9212088.

\bibitem{Slavnov:2003ae}
A.~A. Slavnov.
\newblock Consistent noncommutative quantum gauge theories?
\newblock {\em Phys. Lett.}, B565:246--252, 2003, hep-th/0304141.

\bibitem{Sochichiu:2000kz}
Corneliu Sochichiu.
\newblock Exercising in k-theory: Brane condensation without tachyon.
\newblock 2000, hep-th/0012262.

\bibitem{Sochichiu:2000ud}
Corneliu Sochichiu.
\newblock M(any) vacua of iib.
\newblock {\em JHEP}, 05:026, 2000, hep-th/0004062.

\bibitem{Sochichiu:2000fs}
Corneliu Sochichiu.
\newblock Matrix models: Fermion doubling vs. anomaly.
\newblock {\em Phys. Lett.}, B485:202--207, 2000, hep-th/0005156.

\bibitem{Sochichiu:2000kr}
Corneliu Sochichiu.
\newblock A note on noncommutative and false noncommutative spaces.
\newblock 2000, hep-th/0010149.

\bibitem{Sochichiu:2000bg}
Corneliu Sochichiu.
\newblock On the equivalence of noncommutative models in various dimensions.
\newblock {\em JHEP}, 08:048, 2000, hep-th/0007127.

\bibitem{Sochichiu:2002jh}
Corneliu Sochichiu.
\newblock Gauge invariance and noncommutativity.
\newblock 2002, hep-th/0202014.

\bibitem{Sochichiu:2002ta}
Corneliu Sochichiu.
\newblock Continuum limit(s) of bmn matrix model: Where is the (nonabelian)
  gauge group?
\newblock {\em Phys. Lett.}, B574:105--110, 2003, hep-th/0206239.

\bibitem{Valtancoli:2002rx}
Paolo Valtancoli.
\newblock Stability of the fuzzy sphere solution from matrix model.
\newblock {\em Int. J. Mod. Phys.}, A18:967, 2003, hep-th/0206075.

\bibitem{VanRaamsdonk:2000rr}
Mark Van~Raamsdonk and Nathan Seiberg.
\newblock Comments on noncommutative perturbative dynamics.
\newblock {\em JHEP}, 03:035, 2000, hep-th/0002186.

\bibitem{Verbaarschot:2005rj}
J.~J.~M. Verbaarschot.
\newblock Qcd, chiral random matrix theory and integrability.
\newblock 2005, hep-th/0502029.

\bibitem{Verbaarschot:1994qf}
Jacobus J.~M. Verbaarschot.
\newblock The spectrum of the qcd dirac operator and chiral random matrix
  theory: The threefold way.
\newblock {\em Phys. Rev. Lett.}, 72:2531--2533, 1994, hep-th/9401059.

\end{thebibliography}
%
% Non-BibTeX users please follow the syntax
% the syntax of "referenc.tex" for your own citations
% \input{referenc}
%%%%%%%%%%%%%%%%%%%%%%%%%%%%%%%%%%%%%%%%%%%%%%%%%%%%%%%%%%%%%%%%%%%%%%  }

%%%%%%%%%%%%%%%%%%%%%%%%%%%%%%%%%%%%%%%%%%%%%%%%%%%%%%%%%%%%%%%%%%%%%%

\printindex
\end{document}